\documentclass[sigconf,nonacm]{acmart}
\AtBeginDocument{%
  }

\acmSubmissionID{4507}

\usepackage{graphicx}
\usepackage{makecell}
\usepackage{multirow}
\usepackage{enumitem}
\usepackage{subfig}

\definecolor{darkgreen}{RGB}{30,150,30}
\definecolor{darkblue}{RGB}{0,0,127}
\definecolor{darkyellow}{RGB}{171,133,0}
\definecolor{darkred}{RGB}{180,20,20}
\definecolor{darkmagenta}{RGB}{127,0,127}
\definecolor{darkcyan}{RGB}{0,127,127}
\definecolor{orange}{RGB}{255,165,0}

\makeatletter
\renewcommand\@copyrightpermission{}
\renewcommand\@copyrightowner{}
\renewcommand\@formatdoi[1]{}
\makeatother
\setcopyright{none}
\acmConference{}{}{}
\acmBooktitle{}
\acmYear{}
\acmISBN{}
\acmDOI{}
\begin{document}

\title{SiCo: An Interactive \underline{Si}ze-\underline{Co}ntrollable Virtual Try-On Approach for Informed Decision-Making}

\author{Sherry X. Chen}
\orcid{0000-0002-4964-5286}
\affiliation{%
  \institution{University of California}
  \city{Santa Barbara}
  \country{USA}
}
\email{xchen774@ucsb.edu}

\author{Alex Christopher Lim}
\orcid{0009-0001-3129-4668}
\affiliation{%
  \institution{University of California}
  \city{Santa Barbara}
  \country{USA}
}
\email{xlm@google.com}

\author{Yimeng Liu}
\orcid{0000-0002-6742-2908}
\affiliation{%
  \institution{University of California}
  \city{Santa Barbara}
  \country{USA}
}
\email{yimengliu@ucsb.edu}

\author{Pradeep Sen}
\orcid{0000-0002-8042-924X}
\affiliation{%
  \institution{University of California}
  \city{Santa Barbara}
  \country{USA}
}
\email{psen@ucsb.edu}

\author{Misha Sra}
\orcid{0000-0001-8154-8518}
\affiliation{%
  \institution{University of California}
  \city{Santa Barbara}
  \country{USA}
}
\email{sra@ucsb.edu}

\renewcommand{\shortauthors}{Chen et al.}

\begin{teaserfigure}
    \centering
    \includegraphics[width=\linewidth]{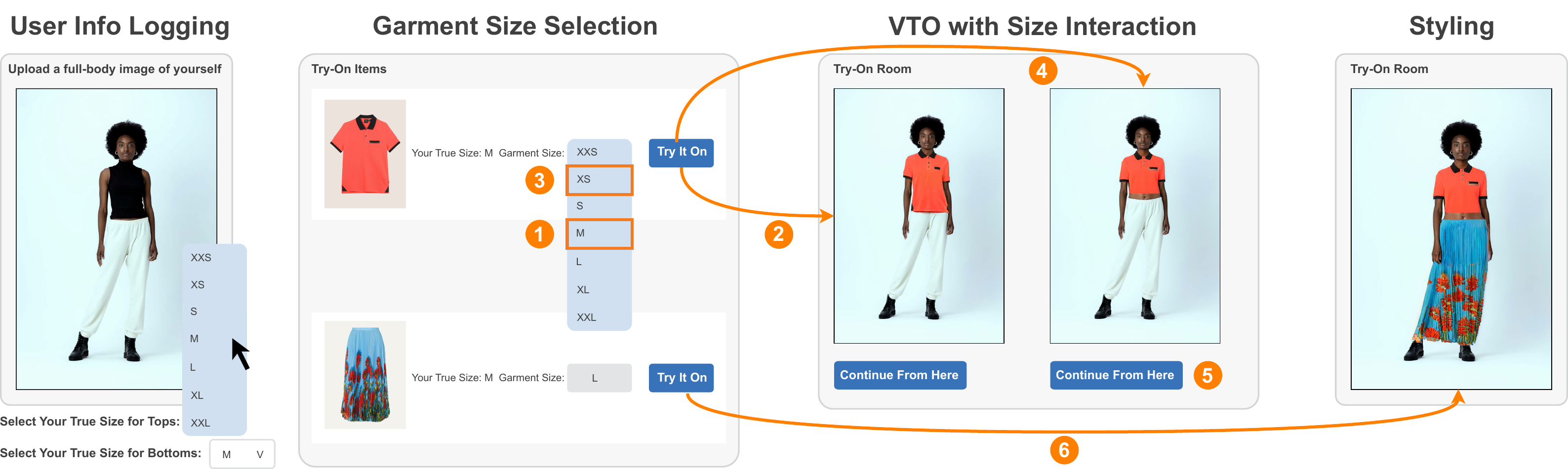}
    \caption{SiCo Overview. Our system begins by prompting users to upload an image of themselves and provide their actual sizes for tops and bottoms. Users then proceed to the garment selection step, where they can choose different sizes for an item (Steps 1 and 3) and visualize how it would look on them (Steps 2 and 4, respectively). In addition to trying on items individually, users can also style multiple items together (Steps 5 and 6). The features in our system enable users to interact with garment sizes and fits more intuitively, providing valuable insights to help them make informed decisions about which garments and sizes to purchase. Garment images are sourced from the DressCode dataset~\cite{morelli2022dress}. Model image \textcopyright\ Pexels.}
    \label{fig:teaser}
    \Description{User workflow of using our virtual try-on system, including user info logging, garment size selection, virtual try-on with size interaction, and styling.}
\end{teaserfigure}

\begin{abstract}
Virtual try-on (VTO) applications aim to replicate the in-store shopping experience and enhance online shopping by enabling users to interact with garments. However, many existing tools adopt a one-size-fits-all approach when visualizing clothing items. This approach limits user interaction with garments, particularly regarding size and fit adjustments, and fails to provide direct insights for size recommendations. As a result, these limitations contribute to high return rates in online shopping. To address this, we introduce SiCo, a new online VTO system that allows users to upload images of themselves and interact with garments by visualizing how different sizes would fit their bodies. Our user study demonstrates that our approach significantly improves users’ ability to assess how outfits will appear on their bodies and increases their confidence in selecting clothing sizes that align with their preferences. Based on our evaluation, we believe that SiCo has the potential to reduce return rates and transform the online clothing shopping experience.
\end{abstract}





\maketitle

\section{Introduction\label{sec:introduction}}

% \begin{figure}[b]
%     \centering
%     \includegraphics[width=0.8\linewidth]{fig/1_introduction/google_vto.pdf}
    
%     \includegraphics[width=0.9\linewidth]{fig/1_introduction/walmart_vto.pdf}
%     \caption{Virtual Try-On Services by Google (top) and Walmart (bottom). While Google and Walmart support virtual try-on with models of different sizes, customers can only visualize how a garment in its regular fit would look, without the option to explore different fits of the same clothing.}
%     \label{fig:google_walmart}
%     \Description{Google and Walmart virtual try-on mobile app that do not allow users to interactive with garment fit.}
% \end{figure}

Online clothing shopping has seen a marked rise in popularity in recent years. According to a 2024 survey by ConsumerX~\cite{consumerx}, 34.1\% of 7,714 respondents aged 15–94 across 14 countries reported shopping online at least once a week, with only 3.1\% having no prior online shopping experience. Among those who shop online, fashion apparel and accessories constitute the largest category, accounting for 56\% of purchases. This growing trend can be attributed to a combination of factors, including convenience (e.g., free shipping, time savings, 24/7 access), financial incentives (e.g., discounts, promotions, lower overall prices), and accessibility.

Online apparel platforms are especially beneficial for people who may face barriers in accessing physical stores or finding appropriate clothing options such as adaptive apparel~\cite{rana2024adaptive} such as those with disabilities, limited geographic access, or lack of transportation. Online environments often offer more detailed product descriptions, which can help individuals with sensory sensitivities such as those with autism spectrum disorder select materials that minimize discomfort~\cite{kay2024sensory, kyriacou2023clothes}. Online communities can also provide supportive spaces for exchanging information and advice regarding clothing items~\cite{annett2015shopping}. Additionally, cultural norms and social media influence body image perceptions, making online shopping a more comfortable alternative for individuals who prefer to avoid in-person experiences due to body image concerns~\cite{merino2024body}. 
 
However, most websites only provide static garment images and model try-on photos, offering users limited ways to interact with products. Virtual try-on (VTO) applications have been developed to bridge the experiential gap by providing enhanced visual information to shoppers~\cite{merle2012whether, zhang2019role, bialkova2022virtual}. These solutions leverage 2D screen-based interfaces as well as 3D technologies, including augmented reality (AR) and virtual reality (VR) methods~\cite{liu2020comparing}. A number of prior works have shown that VTO can significantly enhance product visualization and offer greater sensory and emotional engagement by providing a personalized experience~\cite{chen2024virtual, wang2024study, kim2024unveiling}. 

Critically, VTO also addresses one of the most pressing issues in online shopping: high return rates. Compared to in-store shopping, online purchases are returned more frequently, leading to increased waste and environmental impact~\cite{shopify, chainstoreage}. By helping consumers better evaluate how garments might appear on their bodies, VTO can increase purchase confidence and reduce the likelihood of returns~\cite{sekri2024effects}.

Empirical findings support the impact of VTO on return reduction and sustainability. For instance, a recent study by Vogue Business~\cite{vogue} reported that VTO experiences significantly reduced return rates, with brands using digital mannequin try-on services observing an average decrease of 25\%. Another 2021 study by Glossy~\cite{glossy} found that brands offering VTO experienced 64\% fewer returns compared to those without it. Real-world examples further validate these findings: Macy's reduced its return rate to less than 2\% after introducing virtual fitting rooms in 2023, while Shopify achieved a 40\% reduction in returns by adopting AR-based VTO solutions.

Despite these advances, size and fit issues remain largely overlooked, though they account for 53\% of returns in the U.S. according to a 2023 Coresight Research survey~\cite{coresight}. Most VTO methods adopt a one‑size‑fits‑all approach, assuming garments fit all bodies uniformly. This prevents users from adjusting garment fit in visualizations, ultimately failing to show how clothing will truly fit their unique body shapes. 
% (Fig.~\ref{fig:google_walmart})

To address these limitations, in this work we focus on image‑based VTO methods, as they offer significant potential for seamless integration into existing online shopping platforms. Unlike approaches that require additional hardware such as VR/AR headsets~\cite{yuan2011mixed, liu2020comparing}, complex setups~\cite{giovanni2012virtual, yuan2013mixed}, or other wearable devices~\cite{chong2021per}, image‑based methods eliminate these barriers, making them more accessible and user‑friendly.

Recent image‑based approaches have made strides in incorporating user measurements, such as torso‑to‑shoulder ratios~\cite{chen2023size}, or simulating how garments fit models of various sizes~\cite{liu2016deepfashion, han2018viton, dong2019towards, choi2021viton, morelli2022dress}. However, these efforts remain limited, failing to fully capture the intricate relationship between a user's actual size and garment dimensions or to accommodate individual preferences, such as tighter or looser fits. Furthermore, the machine learning models powering these methods are often trained on datasets featuring predominantly slim body types~\cite{liu2016deepfashion, han2018viton, dong2019towards, choi2021viton, morelli2022dress}, introducing biases that distort user identity and exclude a broader range of body shapes from VTO visualizations. By empowering users to tailor garments in image‑based VTO visualizations to their unique body measurements, retailers could potentially reduce the 53\% return rate stemming from size and fit issues.

%In this work, we focus on image-based VTO methods, as they offer significant potential for seamless integration into existing online shopping platforms. Unlike approaches that require additional hardware, such as VR/AR headsets~\cite{yuan2011mixed, liu2020comparing}, complex setups~\cite{giovanni2012virtual, yuan2013mixed}, or other wearable devices~\cite{chong2021per}, image-based methods eliminate these barriers, making them more accessible and user-friendly.

%Recent image-based approaches have made strides in incorporating user measurements, such as torso-to-shoulder ratios~\cite{chen2023size}, or simulating how garments fit models of various sizes~\cite{googlevto} (Fig.\ref{fig:google_walmart}). However, these efforts remain limited, failing to fully capture the intricate relationship between a user's actual size and garment dimensions or to accommodate individual preferences, such as tighter or looser fits. Furthermore, the machine learning models powering these methods are often trained on datasets featuring predominantly slim body types~\cite{liu2016deepfashion, han2018viton, dong2019towards, choi2021viton, morelli2022dress}, introducing biases that distort user identity and exclude a broader range of body shapes from VTO visualizations.

To this end, we present SiCo, an interactive VTO system that enables users to explore how garments in various sizes would look on them (Fig.~\ref{fig:teaser}). Users can input their information, including a self-image and their actual sizes for tops and bottoms, or select from a set of provided model images to preserve privacy while viewing garments on a body shape similar to their own. The interface is designed to resemble popular online clothing stores, offering a familiar and comfortable user experience while integrating features that enhance user ability to interact with garment sizes. Users can experiment with different sizes of the same garment, compare VTO visualizations side by side, and style multiple garments together to visualize complete outfits.

The generative backbone of our system combines Stable Diffusion~\cite{rombach2022high} and IP-Adapter~\cite{ye2023ip}, augmented with a garment-sizing mechanism that we designed to mitigate model bias. This approach ensures the system avoids generating unrealistically slim or muscular figures, addressing biases commonly found in training datasets. Unlike commercial services from companies like Google\footnote{https://blog.google/products/shopping/ai-virtual-try-on-google-shopping} and Walmart\footnote{https://www.walmart.com/cp/virtual-try-on/4556767}, which only support regular-fit visualizations on models of varying sizes, SiCo offers a more personalized and user-centric experience offering additional options such as loose or tight fits. By addressing critical size and fit challenges in online clothing shopping, our system empowers users to make more informed purchasing decisions, enhancing the overall shopping experience.

To evaluate the impact of SiCo on improving decision-making in online clothing shopping, we conducted a user study with 48 participants. Each participant interacted with two out of four system versions: some included size-controllable functionality, while others provided baseline VTO features. The results revealed a strong preference for the size-controllable version. Participants reported that this feature significantly improved their ability to visualize how outfits would look on them, enhanced their understanding of garment fit and appearance, and increased their confidence in making clothing choices.

\section{Related Work}

\subsection{VTO applications}

VTO has become a hot topic in the retail industry due to its vast commercial potential. This technology allows consumers to virtually see themselves in different clothing items without requiring a physical try-on, offering a more personalized shopping experience through tailored clothing recommendations based on a customer's body type, style preferences, and past purchases.

While many prior works have focused on VTO for clothing try-ons, the technology has also been extended to other items, such as glasses, shoes, and accessories. For instance, Liu et al.~\cite{liu2020comparing} compared VTO using personalized animated avatars in XR (Extended Reality) with traditional online interfaces, showing that XR positively influences the shopping experience by enabling realistic garment visualization. Yuan et al.~\cite{yuan2013mixed} developed an XR application for VTO that lets users view virtual clothing from different angles in real-time as they move. Similarly, Giovanni et al.~\cite{giovanni2012virtual} utilized a Kinect sensor and a High-Definition (HD) camera to enhance the performance of VTO in XR. Recently, VTO has been integrated into mobile AR (Augmented Reality) applications like Snapchat, making VTO accessible to a broader audience through smartphones~\cite{snapchat_vto}.
Beyond XR techniques, Chong et al.~\cite{chong2021per} designed an actuated mannequin to capture garment deformations under various body poses to help enhance the realism of VTOs. 

While these applications have advanced the VTO experience using immersive and spatial technologies, most require complex and precise hardware setups, including AR/VR headsets, wearable devices, and data capture sensors. These setups are often time-consuming, expensive, and difficult to integrate into existing at-home online shopping experiences.

Our system addresses these limitations by building on familiar web technologies, eliminating the need for additional hardware. This approach ensures a more accessible and seamless VTO experience for users, aligning with the convenience of current online shopping platforms.

\subsection{Image-based VTO methods}

VTO methods aim to create realistic depictions of individuals wearing selected garments. Image-based VTO techniques focus specifically on achieving this with 2D images by superimposing clothing onto human figures. Traditional VTO approaches begin by parsing human images to extract markers, poses, and dense pose data using pre-trained models~\cite{cao2017realtime,guler2018densepose}. This information is then used to align target garments with the human body~\cite{han2018viton,wang2018toward}. Many studies have adopted this approach, parsing input images to generate various representations, such as human-clothing segmentation~\cite{li2021toward,choi2021viton} and clothing-agnostic person representations~\cite{han2019clothflow,yu2019vtnfp,ge2021disentangled}.
In contrast, another line of research~\cite{issenhuth2020not,ge2021parser,he2022style} highlights the limitations of methods that rely heavily on external pre-trained parsers. These studies advocate for a parser-free strategy, where models independently learn intrinsic human representations for garment wrapping, which are derived either from parsing-based methods~\cite{ge2021parser} or pre-trained garment-wrapping modules~\cite{issenhuth2020not}.

Despite these innovations, the limited capacity of these methods' architectures and biases in training datasets often compromise their quality and ability to be generalized to different use cases. Recent advancements~\cite{choi2024improving,gou2023taming,zhu2023tryondiffusion,kim2023stableviton} have leveraged diffusion models~\cite{dhariwal2021diffusion,song2020denoising,rombach2022high} instead of traditional GANs~\cite{goodfellow2020generative}, enabling the development of higher-quality and more robust applications. Some approaches continue to incorporate parsed information, such as human pose and segmentation, to guide the generation process~\cite{choi2024improving,gou2023taming}, while others rely solely on human-garment image pairs for the model to synthesize high-quality results~\cite{zhu2023tryondiffusion,kim2023stableviton}.

However, many of these newer methods assume a generalized fit for garments across all body types, significantly limiting their effectiveness in helping customers make critical decisions about garment sizing. For example, Google’s virtual try-on feature supports models of different sizes but only displays a standard fit, neglecting alternative fits like oversized or cropped options that users may prefer. Furthermore, existing studies that incorporate specific measurements, such as shoulder-to-shoulder ratios~\cite{chen2023size}, often fail to provide intuitive and actionable garment size recommendations that facilitate decision-making.

Generative models such as Stable Diffusion (SD)~\cite{rombach2022high} can produce a wide range of images but are not immune to biases, including those related to race and gender~\cite{anivcin2022bias,chauhan2024identifying,wu2023stable}. These biases can compromise the authenticity of VTO outcomes by distorting user identity. Our experiments reveal that SD tends to alter user body shapes, making them appear slimmer or more muscular, which reduces the authenticity of the try-on experience and limits its utility in VTO system design.

To address these issues, our system employs an identity preservation mechanism that conditions SD outputs on the user’s body contour, ensuring the user's body shape remains accurate. This approach reduces biases inherent in AI models and improves inclusive-ness, leading to a better online shopping experience for a more diverse range of customers.

\section{Design Objectives}

Our work addresses the limitations of existing image-based VTO systems, which fail to enable users to manipulate garment sizes and fits. Our objective is to enhance user interactivity with garment sizing and fitting, to create a user-friendly solution that provides intuitive insights into how clothing of various sizes would fit a specific user's body type. To achieve this, we established the following design objectives.

\vspace{0.1in}\noindent\textbf{Integration into current user shopping experiences (DO1).} Conventional apparel websites limit user ability to interact with clothing, yet they remain popular due to their convenience. To align with this familiarity, we propose leveraging the same medium — websites — for our VTO interface design. While alternative mediums, such as VR/AR~\cite{liu2020comparing, yuan2013mixed, giovanni2012virtual} and wearable devices~\cite{chong2021per}, offer unique advantages, they represent a significant departure from the traditional online shopping experience and are often inaccessible to large segments of the general audience and contradict the low-effort appeal of online shopping. To address these challenges, we designed our VTO system to be entirely web-based, enabling seamless integration with most existing online clothing websites while maintaining convenience and accessibility.

\vspace{0.1in}\noindent\textbf{Simulating the in-store fitting room experience online (DO2).} VTO applications should bridge the gap created by the lack of direct access to clothing, offering visualizations that provide intuitive and accurate information. Traditional websites often rely on users imagining how they would look in garments modeled by individuals with different physiques. While some websites now feature diverse models, this approach still introduces a level of abstraction into the decision-making process. Existing VTO systems that use avatars or 3D human geometry overlays often fail to accurately represent user appearance with these avatars, making it difficult for users to relate to these visualizations. To overcome this limitation, we developed an image-based VTO system. By using photographs, our approach provides accurate visual representations of users, enabling them to relate more easily to the visualizations and make informed purchasing decisions.

\vspace{0.1in}\noindent\textbf{Ease of size indication (DO3).} Size indication is a critical yet challenging aspect of online clothing purchases. Many websites provide product measurements, requiring users to compare these with their own measurements or similar items they own. Others rely on size-recommendation tools that require extensive personal information, such as height, weight, age, bra size, and waist size, before offering suggestions. However, users are often unsure of their measurements, and obtaining accurate measurements on the spot can be difficult. Instead, our system requests a user's true size in the form of standard size labels, ranging from XXS to XXL. These labels are familiar, easily accessible, and directly aligned with the garment size options available during the purchasing process.

\section{Interface Design}

\subsection{User information logging}

\begin{figure}
    \centering
    \includegraphics[trim=0 60 0 0, clip, width=0.9\linewidth]{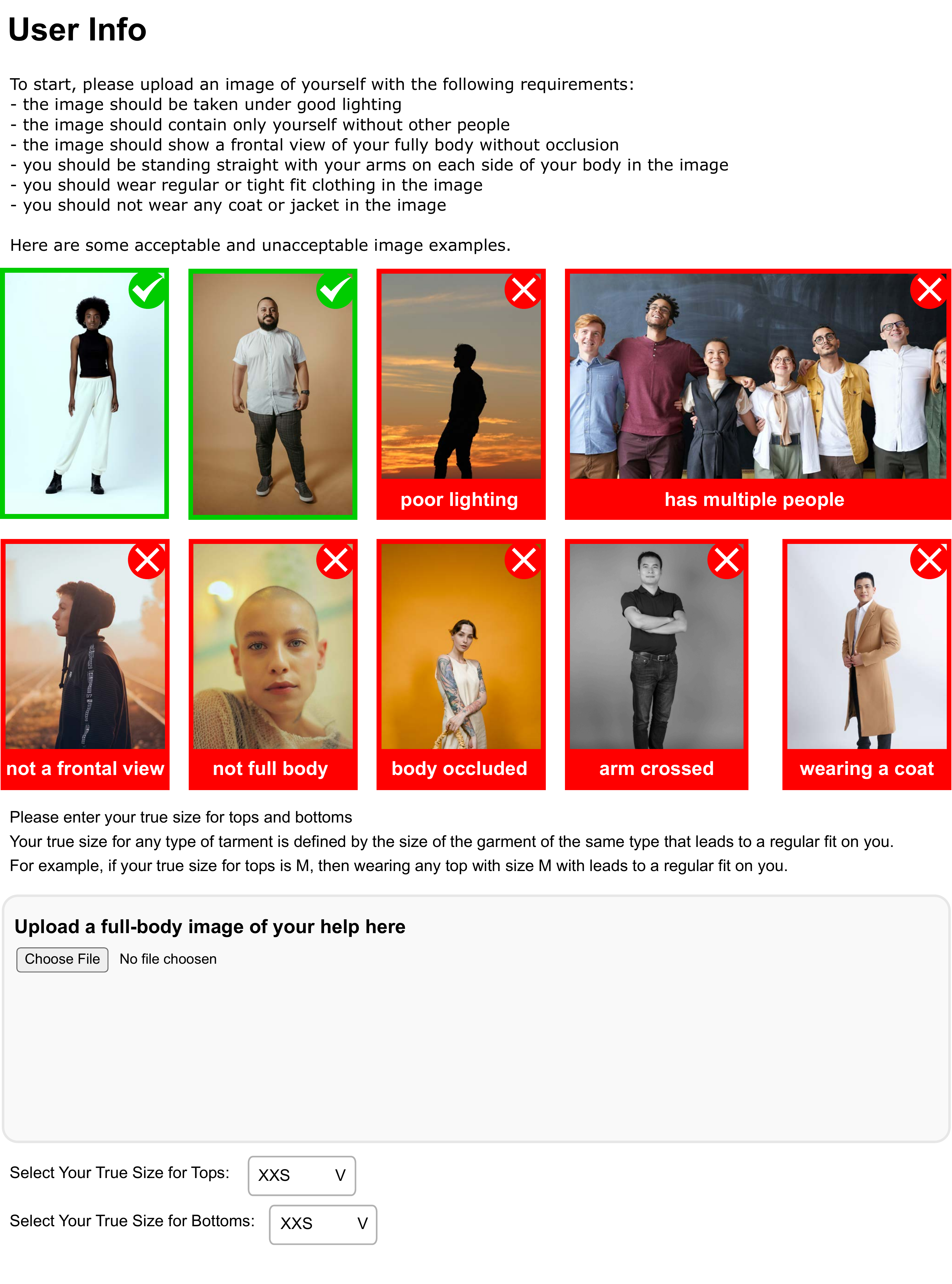}
    \caption{User Info Logging. The user experience begins with a prompt to upload an image of themselves, guided by specific instructions to ensure accurate processing. This step is essential for the system to generate precise and reliable VTO outcomes. Beneath the image upload section, users are asked to input their true sizes (XXS to XL) for tops and bottoms, defined as the garment sizes that provide a regular fit for their body. This information forms the basis for generating VTO visualizations of clothing in various fits, enabling a personalized and accurate try-on experience. Individual image examples in the interface \textcopyright\ Pexels.}
    \label{fig:interface_upload_image}
    \Description{Our user info logging page with examples of suitable and unsuitable user images to upload.}
\end{figure}

\begin{figure}
    \centering
    \includegraphics[width=0.9\linewidth]{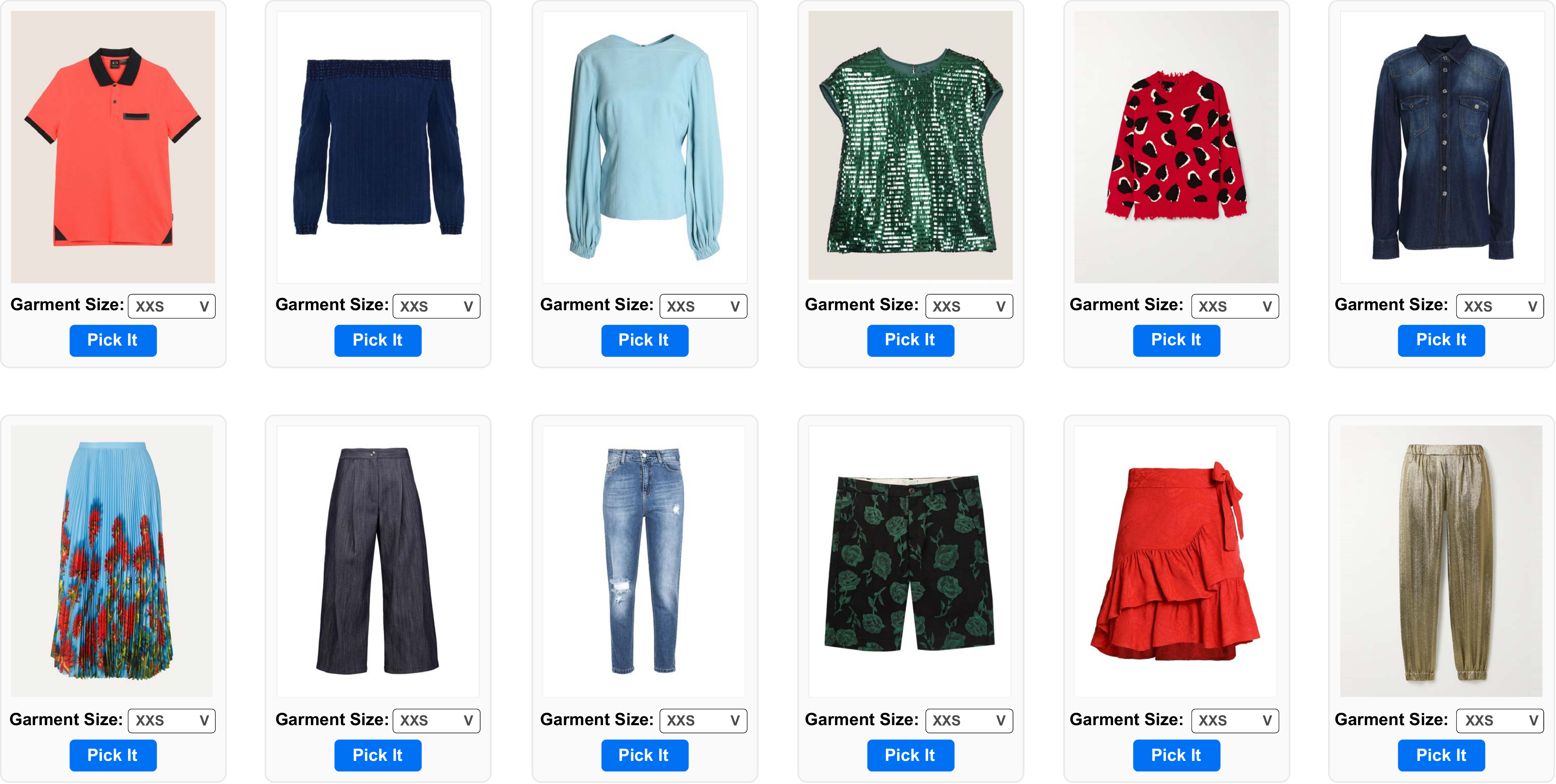}
    \caption{Garment Selection. Our interface offers a curated collection of garments, each accompanied by size selection options tailored to the user's true-size input. The design closely mirrors conventional clothing websites, providing a familiar and intuitive shopping experience. Garment images are sourced from the Dress Code dataset~\cite{morelli2022dress}.}
    \label{fig:interface_with_size}
    \Description{Our garment selection page with clothing items and sizes for user to choose.}
\end{figure}

\begin{figure*}
    \centering
    \includegraphics[width=0.95\linewidth]{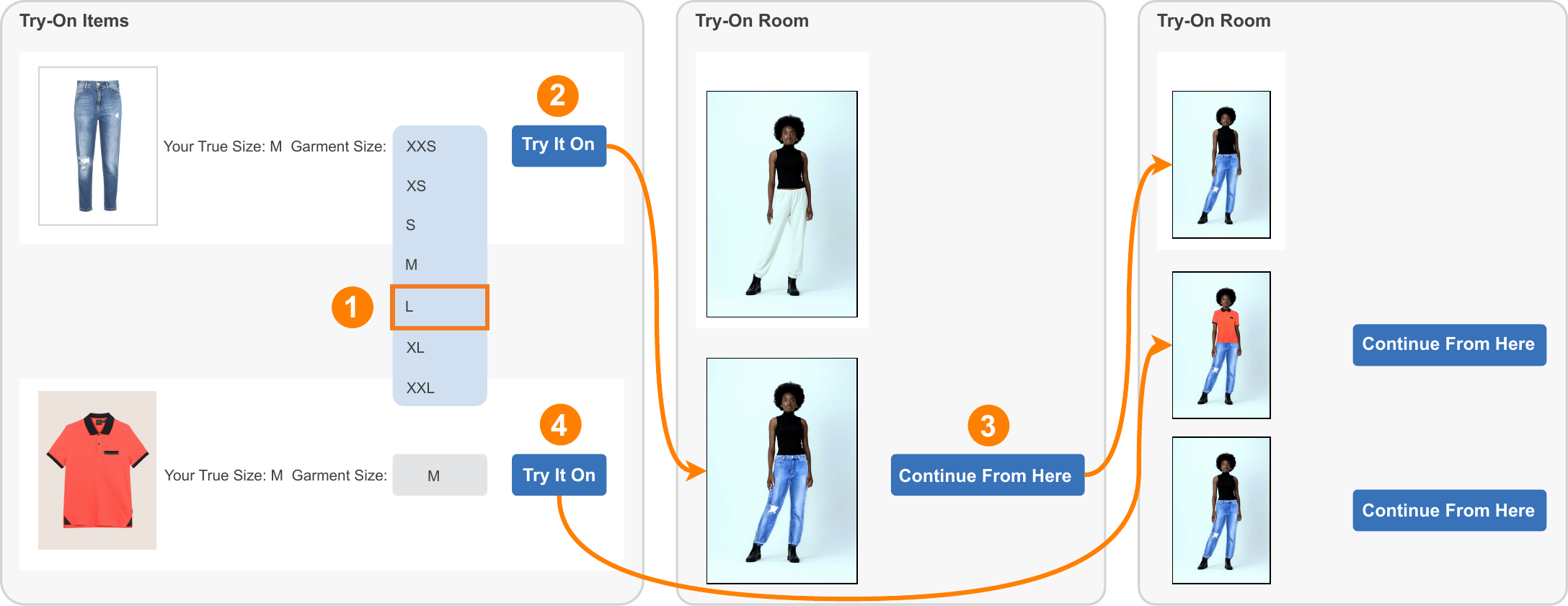}
    \caption{Users can experiment with their selected garments in the ``Try-On Items'' section, where each item is accompanied by the user's true size (specified at the start of their session) and the garment size, which can be adjusted at any time. The self-image uploaded by the user at the beginning appears in the ``Try-On Room'' section under ``Before Try-On,'' serving as the foundation for all subsequent virtual try-on (VTO) visualizations. To try on a garment, users simply select their preferred size and click the ``Try It On'' button to generate the corresponding visualization (Steps 1 and 2). To facilitate multi-garment styling, each visualization includes a ``Continue From Here'' button. By clicking this button, users can update the ``Before Try-On'' image to reflect the current visualization (Step 3), enabling the layering and styling of multiple garments (Step 4). Garment images are sourced from the Dress Code dataset~\cite{morelli2022dress}. Model image \textcopyright\ Pexels.}
    \label{fig:interface_try_on}
    \Description{User workflow of using our virtual try-on system.}
\end{figure*}

We designed a web-based interface to guide users through multiple stages, each replicating aspects of the in-store shopping experience (Fig.~\ref{fig:teaser}). Our primary focus is on scenarios where users are shopping for themselves. In a physical store, users rely on their bodies as a reference to select garment sizes and fits before making a purchase. To emulate this process in our VTO system, we ensure users log essential information at the outset.

Much like Warby Parker\footnote{https://www.warbyparker.com/app} asks a user to upload a selfie to virtually try on frames or apps such as YouCam Makeup\footnote{https://www.perfectcorp.com/consumer/apps/ymk} or L'Oréal's Style My Hair\footnote{https://www.loreal.com/en/articles/science-and-technology/haircolor-virtual-try-on-loreal-professionnel-style-myhair/} let a user preview new looks or fitness services such as Freeletics\footnote{https://www.freeletics.com/en/} track pre- and post-workout body changes with user self-images, our VTO system also employs user‑uploaded self-images to generate customized clothing size and fit visualizations. As shown in Fig.~\ref{fig:interface_upload_image}, users are prompted to upload an image of themselves standing against a plain background, enabling the system to accurately process their proportions and deliver highly customized try‑on results. To accommodate users with privacy concerns or discomfort about uploading personal photos, we also provide an option to select from a curated set of model images representing various body types that allow everyone to explore fit and style VTO while maintaining their comfort and privacy.

%Specifically, as shown in Fig.~\ref{fig:interface_upload_image}, users are prompted to upload an image of themselves standing against a plain background. This step enables the system to accurately process the uploaded images, resulting in precise visualizations. To accommodate users with privacy concerns or discomfort about uploading personal images, we also provide the option for users to select from a set of model images representing different body types. This alternative allows users to engage with the system while maintaining their privacy and comfort.

Below the image upload section, users are asked to input their true sizes for tops and bottoms. These true sizes, which are defined as garment sizes that provide a regular fit for their body, are selected from seven standardized labels: XXS, XS, S, M, L, XL, and XXL. This information forms the basis for generating VTO visualizations of clothing with various fits, enhancing the personalization and accuracy of the user experience.

\begin{figure*}
\centering
\subfloat[Overview]{\includegraphics[width=0.45\linewidth]{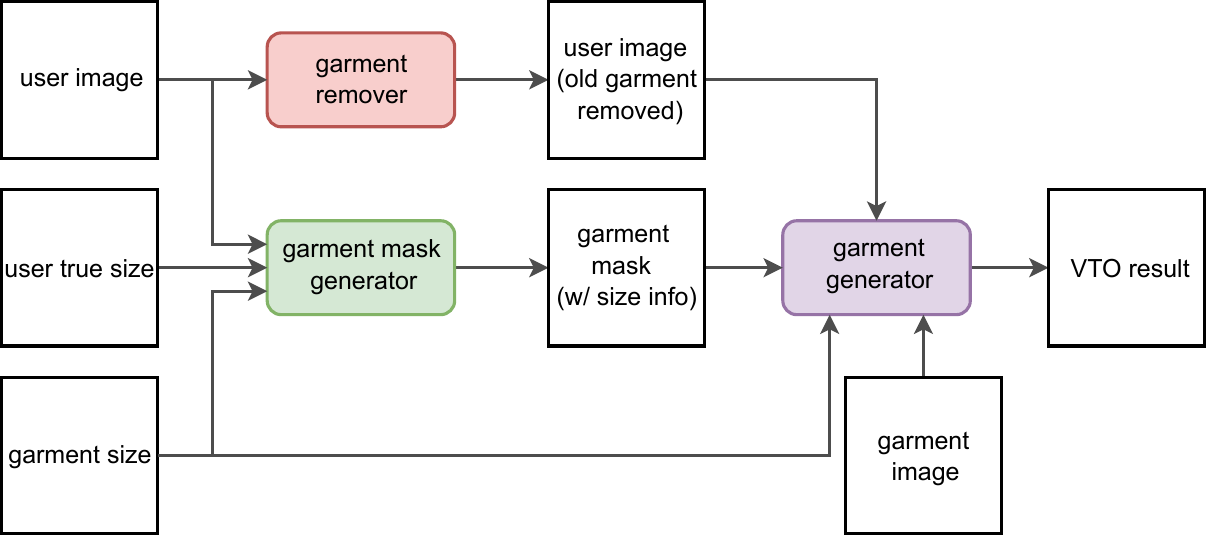}\label{fig:supplementary_system_overview}}
\hspace{15pt}
\subfloat[Garment Remover]{\includegraphics[width=0.45\linewidth]{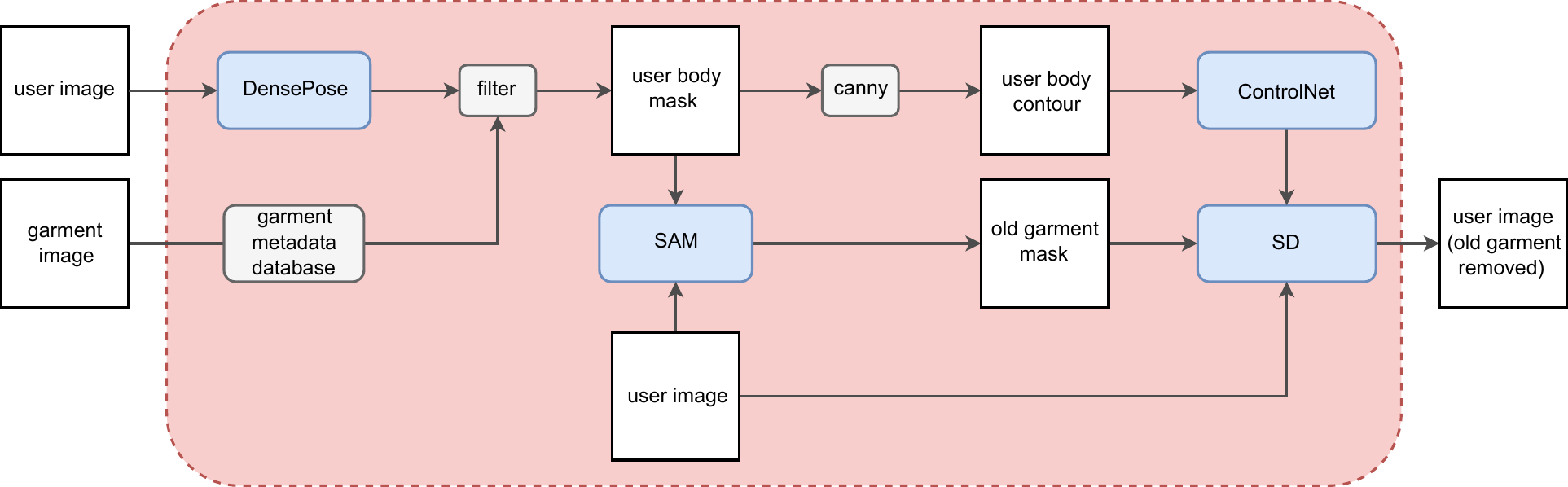}\label{fig:backbone_garment_remover}}

\subfloat[Garment Mask Generator]{\includegraphics[width=0.45\linewidth]{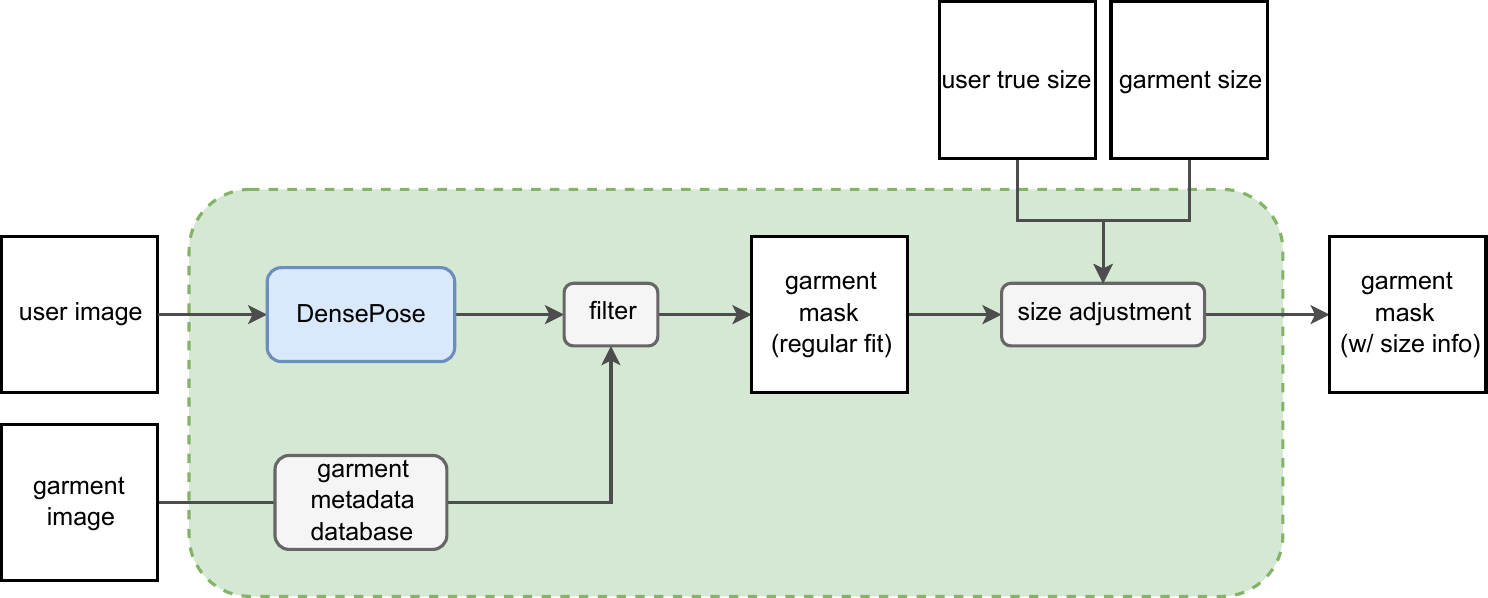}\label{fig:backbone_garment_mask_generator}}
\hspace{15pt}
\subfloat[Garment Generator]{\includegraphics[width=0.45\linewidth]{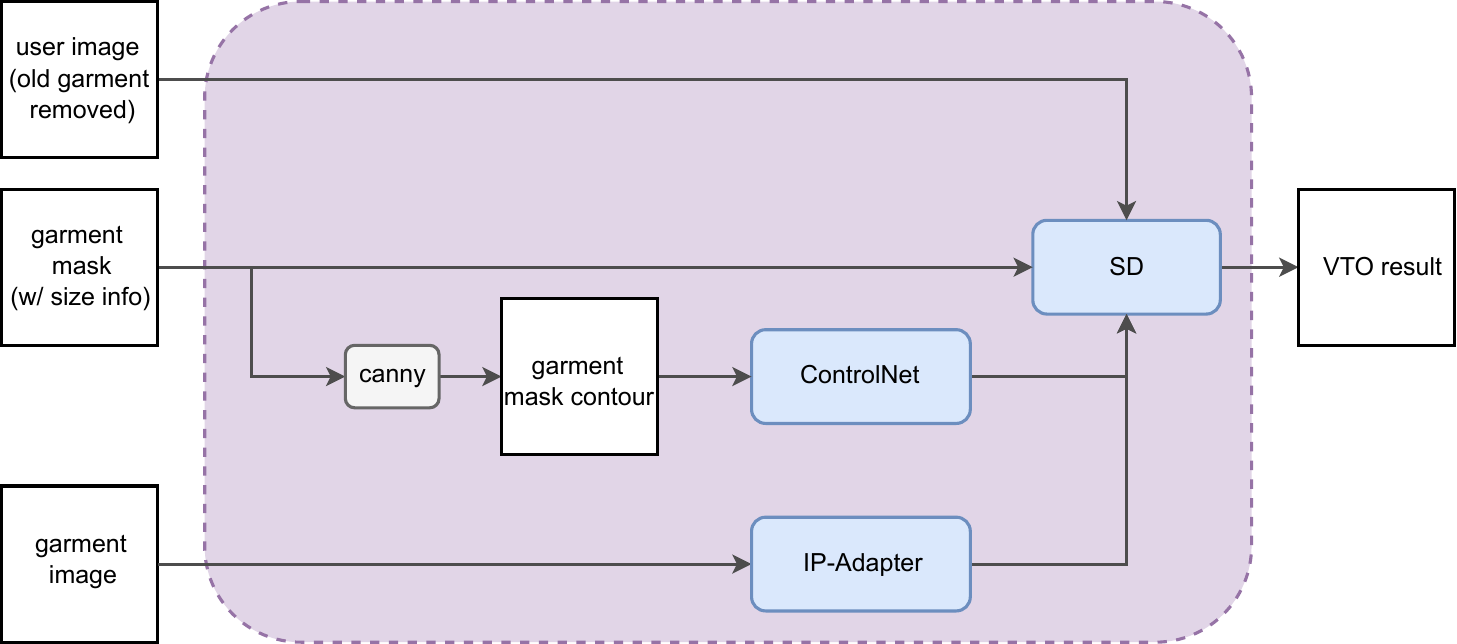}\label{fig:backbone_garment_generator}}
\caption{System Backbone. The backbone comprises an identity-preserved garment remover (Fig.~\ref{fig:backbone_garment_remover}), a size-controllable garment mask generator (Fig.~\ref{fig:backbone_garment_mask_generator}), and a garment generator (Fig.~\ref{fig:backbone_garment_generator}). It processes the user's self-image, true size, selected garment, and size to generate a size-controllable virtual try-on result while preserving the user’s physical identity.}
\label{fig:backbone}
\Description{Flow-charts of our virtual try-on system backbone and its three components: garment remover, garment mask generator, and garment generator.}
\end{figure*}

\subsection{Garment selection}

In the next stage, users select garments of interest (Fig.~\ref{fig:interface_with_size}). This stage features a curated collection of garments with size selection options tailored to the user's true-size input. The interface closely resembles conventional clothing websites, ensuring a familiar and comfortable user experience. As outlined in design objective \textbf{DO1}, most users are more accustomed to traditional online shopping platforms than VTO services, a hypothesis supported by pre-questionnaire statistics from our user study. To ease the transition to interactive garment visualization, we integrate the novel VTO feature into a familiar interface modeled after traditional clothing websites, bridging innovation with familiarity.

Although both user and garment measurements are represented by size labels, this abstraction simplifies size selection for user interaction (\textbf{DO3}). For clothing brands interested in integrating this feature into their websites, detailed measurements for each product can be stored in the system to ensure VTO visualizations align closely with physical try-on results. In our implementation, each garment is paired with pre-computed attributes, such as garment type and sleeve/leg length, which are stored in the system’s database for efficient processing during visualization. This ensures the system delivers accurate and realistic try-on results tailored to the user’s inputs.

\subsection{Try-on with garment size controllability}

The core of the VTO experience unfolds in the next stage, as illustrated in Fig.~\ref{fig:interface_try_on}. During this stage, the garments selected earlier are displayed in the ``Try-On Items'' section, resembling a basket of clothing items users might carry before entering a fitting room or a website's shopping cart.

Each garment is accompanied by two key details: the user's true size, specified at the beginning of their session, and the garment size, which can be adjusted at any time. The true size serves as a reference point, while the adjustable garment size allows users to explore and compare fit differences through side-by-side visualizations. This feature enhances user ability to interact with garment sizes and provides users with a more comprehensive understanding of how each size may look and feel.

The self-image uploaded at the start of the session appears in the ``Try-On Room'' section under ``Before Try-On.'' All subsequent virtual try-on (VTO) visualizations are based on this image. To try on a garment, users select their preferred size and click the ``Try It On'' button. The resulting visualization is displayed in the ``Try-On Room'' section alongside the ``Before Try-On'' image, enabling users to evaluate individual items seamlessly.

For extended visualizations, the ``Continue From Here'' button allows users to update the ``Before Try-On'' image to reflect the current visualization. This functionality supports layering and styling multiple garments together, providing a more comprehensive and engaging try-on experience.

\section{System Backbone Design\label{sec:system_backbone}}

All the functionalities described above are powered by our system backbone (Fig.~\ref{fig:backbone}), which integrates three key components: an \textit{identity-preserved garment remover}, a \textit{size-controllable garment mask generator}, and a \textit{garment generator}. These components ensure that virtual try-on (VTO) visualizations preserve user identity while accurately reflecting the selected garment's size relative to the user. Below, we outline their functionalities.

\subsection{Identity-preserved garment remover}
The garment remover eliminates existing garments from the user’s photo to simulate a new garment overlay (Fig.~\ref{fig:backbone_garment_remover}). Using DensePose~\cite{guler2018densepose}, we estimate old garment locations and refine the mask using the Segment Anything (SAM) model~\cite{kirillov2023segment}. To preserve user identity during inpainting, Stable Diffusion (SD)~\cite{rombach2022high} is guided by a canny-edge ControlNet~\cite{zhang2023adding} using contours derived from the user’s body mask.

\subsection{Size-controllable garment mask generator}
The garment mask generator adjusts the fit of the garment to reflect the user’s true size and the selected garment size (Fig.~\ref{fig:backbone_garment_mask_generator}). Starting with a ``regular fit" mask derived from the user’s body segments, we modify its dimensions based on the size difference between the user and the garment. If the garment size is larger, we dilate the mask; if smaller, we trim it. This ensures a realistic and user-specific fit.

\subsection{Garment generator}
In the final stage, the garment generator overlays the selected garment onto the user’s modified image (Fig.~\ref{fig:backbone_garment_generator}). Stable Diffusion, enhanced by IP-Adapter~\cite{ye2023ip}, uses the processed garment mask and its contour (via a canny-edge ControlNet~\cite{zhang2023adding}) to accurately recreate the garment while preserving the user's identity.

\subsection{Effect of user identity-preservation}
\begin{figure}
\centering
\small
\subfloat[Input]{\includegraphics[width=0.3\linewidth]{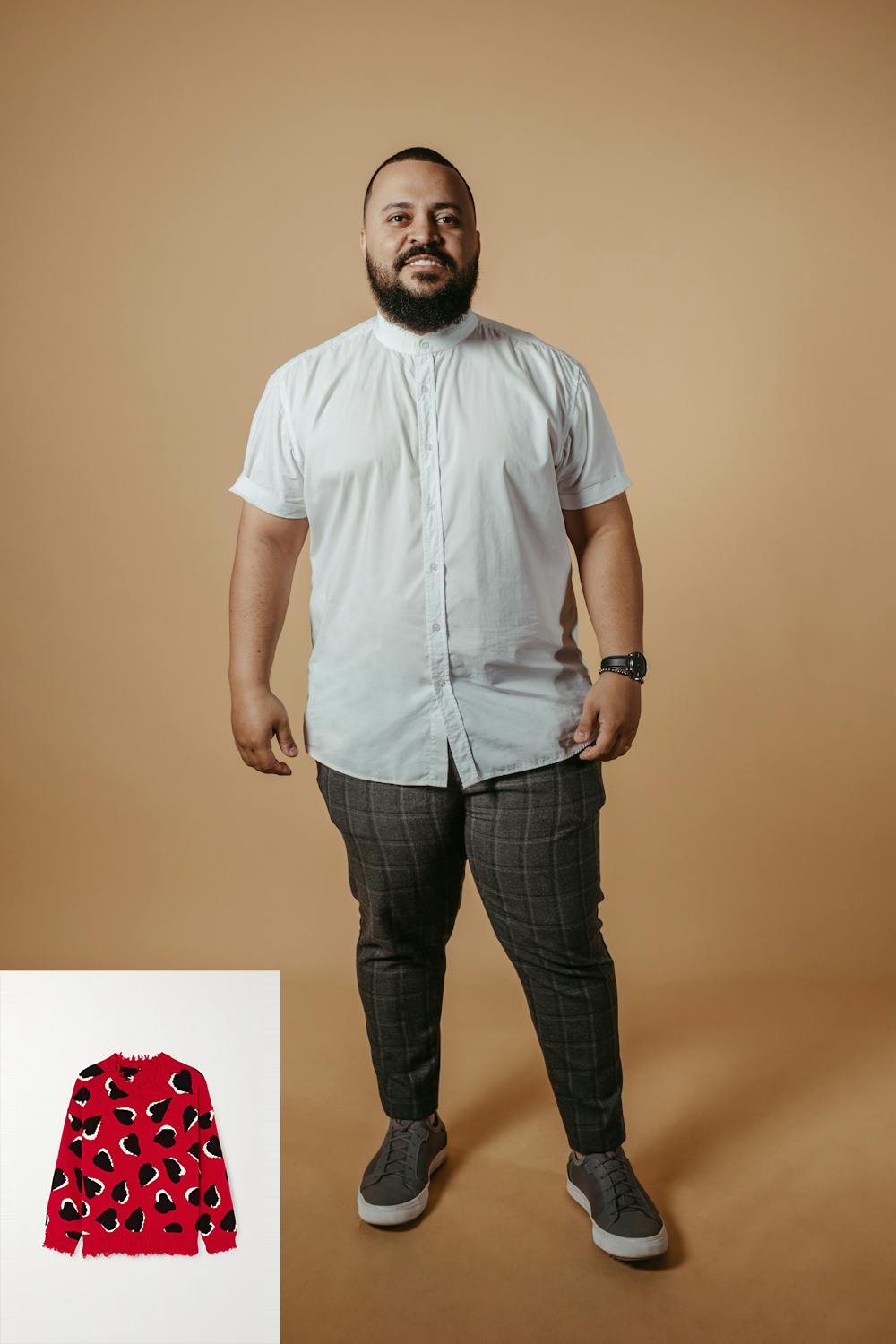}\label{fig:user_garment_input}} 
\hfill
\subfloat[w/o user identity preservation]{\includegraphics[width=0.3\linewidth]{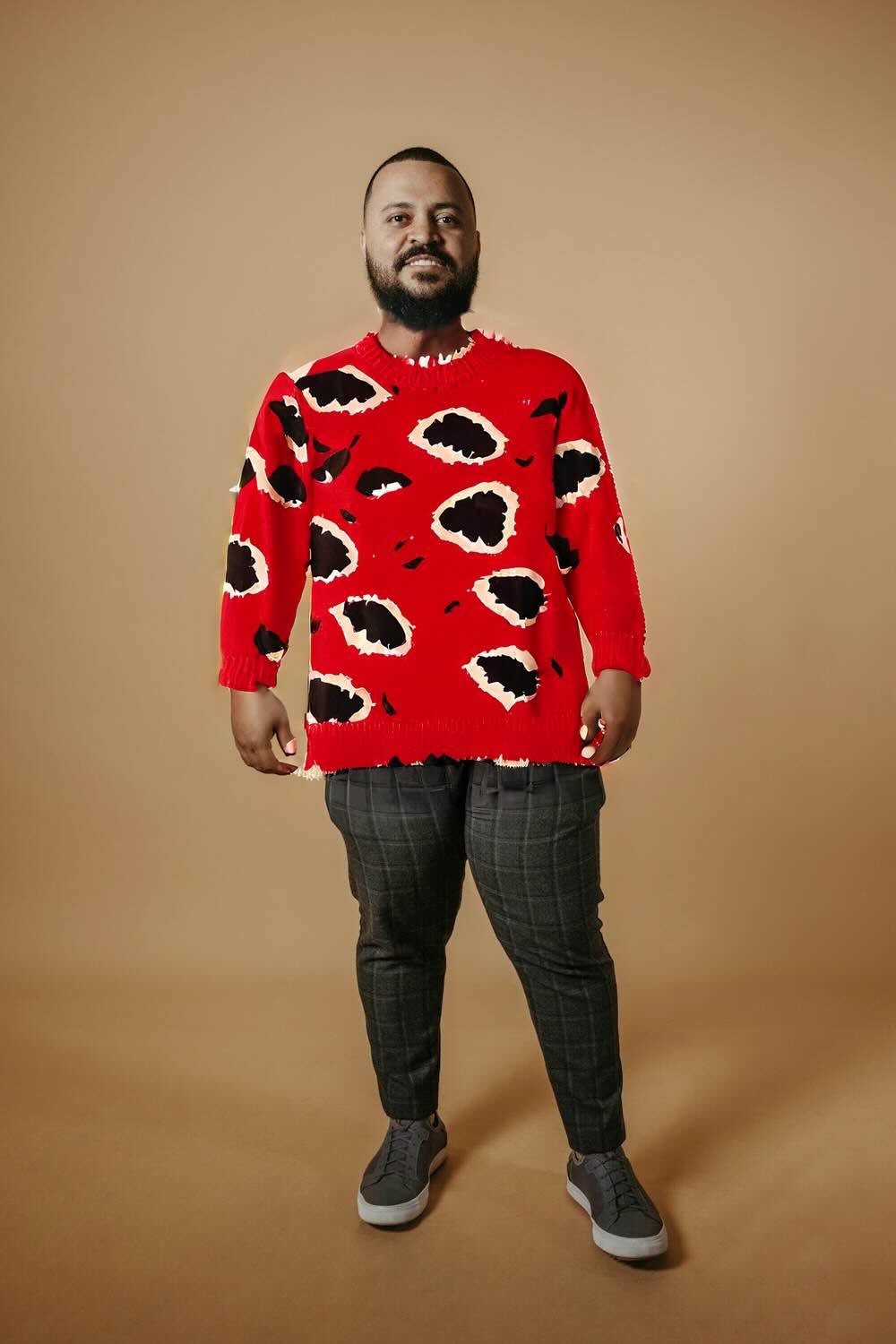}\label{fig:without_contour}}
\hfill
\subfloat[w/ user identity preservation]{\includegraphics[width=0.3\linewidth]{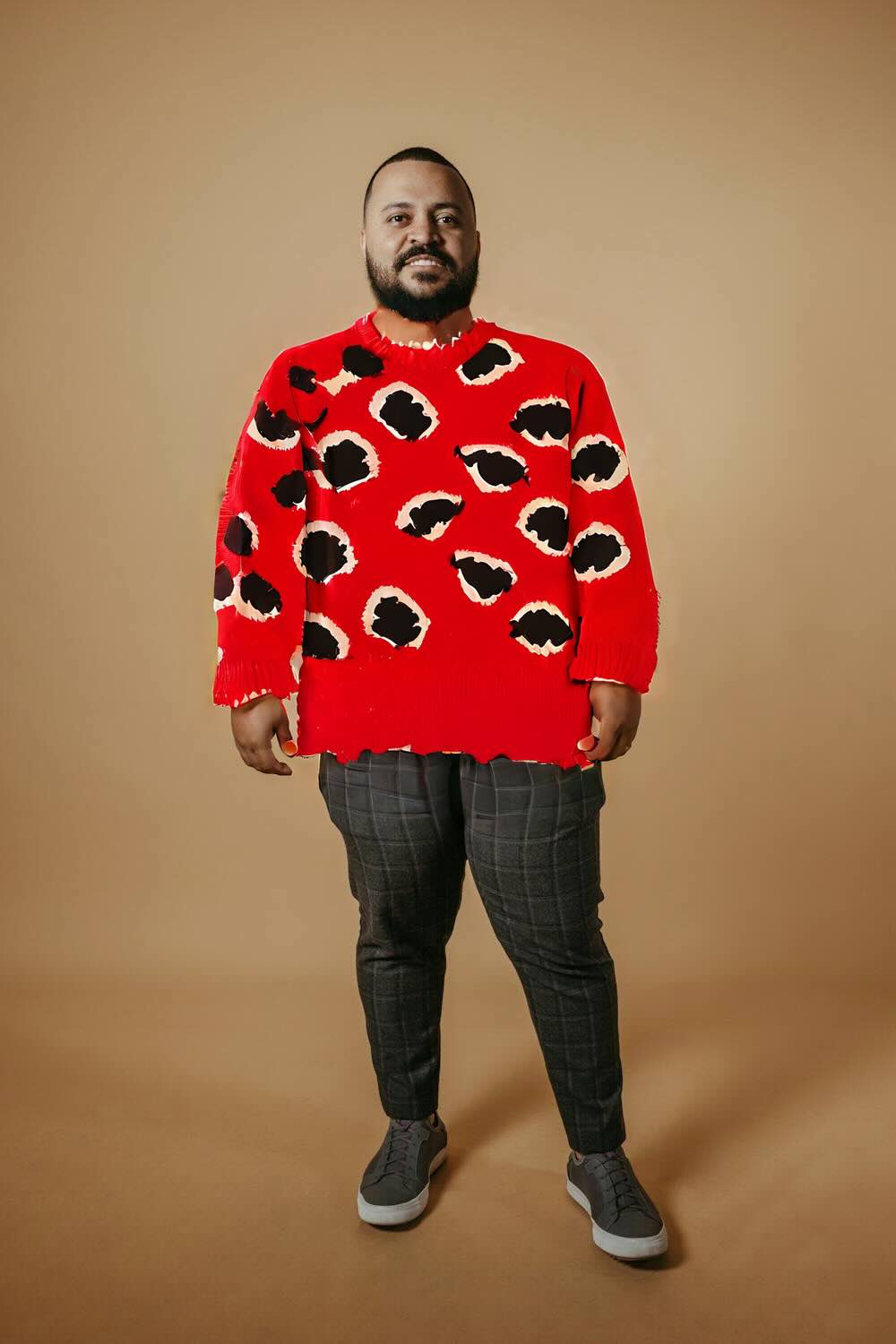}\label{fig:with_contour}}
\caption{Effect of User Identity Preservation. The generative model used as the backbone of our system is Stable Diffusion~\cite{rombach2022high}. While this model is capable of producing high-quality images and is generalizable to various use cases, it tends to generate slim and muscular body types that may not accurately reflect users' physiques due to biases in its training data. Without our user identity preservation mechanism, the system often alters user appearance in virtual try-on (VTO) visualizations (Fig.~\ref{fig:without_contour}). In contrast, our approach prioritizes preserving user identity by maintaining the unique contours and proportions of each user’s body. This mechanism effectively mitigates the inherent biases of Stable Diffusion, ensuring that VTO visualizations remain authentic and representative of the user's true physique (Fig.~\ref{fig:with_contour}). The garment image is sampled from the Dress Code dataset~\cite{morelli2022dress}. The model image \textcopyright Pexels}
\label{fig:body_contour_effect}
\Description{Two side-by-side virtual try-on results of a man trying-on a red sweater to show how our user identity preservation better preserves user physique in the try-on results.}
\end{figure}

A key feature of our backbone design is the user identity preservation mechanism. While Stable Diffusion is utilized for its ability to produce high-quality images and its versatility across various applications, it often generates slim and muscular body types that may not accurately represent end user physiques due to biases in its training data.

Without our identity preservation mechanism, the system frequently modifies user appearances in virtual try-on (VTO) visualizations (Fig.~\ref{fig:without_contour}). In contrast, our approach focuses on maintaining user identity by preserving their unique contours and proportions, effectively mitigating the inherent biases of Stable Diffusion (Fig.~\ref{fig:with_contour}).

\section{Evaluation}

To evaluate SiCo, we recruited participants through various departmental mailing lists on a U.S. university campus which included an overview of the user study with its purpose, study location, duration (30 min), and compensation (\$10). %While this conveninience sampling approach, not unusual for HCI research, limits the diversity of our participant pool, it was chosen to accommodate the requirements of our study, which needed to be conducted in an indoor lab setting, as detailed in Sec.~\ref{sec:apparatus}, where each participant had their photos taken as part of the study, and in-person participation ensured consistent photo quality. Moreover, the controlled environment encouraged participants to engage more seriously with the task. Individuals who volunteered for our study also tended to be more interested and invested, leading to more thoughtful and reliable responses.
Although convenience sampling narrows our participant pool, it was chosen to satisfy our study’s logistical and methodological requirements. Conducting the experiment in an indoor lab setting (see Sec.~\ref{sec:apparatus}) allowed us to capture each participant's photo under consistent lighting and background conditions, and the controlled environment encouraged serious engagement with the task. Moreover, recruiting volunteers ensured that participants were motivated and invested in the study, yielding more thoughtful and reliable responses.

We conducted a preliminary user study with 48 participants, including 27 males, 20 females, and one non-binary individual (average age: 27.2 years). The participants consisted of college students, digital artists, and software engineers. For more information of our participant pool, please refer to the Appendix.

The study aimed to assess how the ability to interact with garment sizing in a Virtual Try-On (VTO) service may influence decision-making in online clothing shopping, compared to a baseline VTO system without this functionality. It also aimed to provide initial insights into cognitive load associated with specifying true and garment sizes, user experience with self-images versus generic model images, and the effort required to prepare self-images compared to using pre-provided model images.

\subsection{Apparatus\label{sec:apparatus}}

Our user study took place in an indoor lab with separate areas for photo capturing and website evaluation. Participants were photographed using an iPhone SE on a tripod two meters from a wall. The website evaluation area featured a computer, keyboard, and mouse for participants to interact with our website.

\subsection{Experiment Design}

\begin{table}[]
    \centering
    \begin{tabular}{c|ccc} 
    \toprule
         & size ctrl? & self-image? & count  \\ \midrule
    A     &  \textcolor{darkgreen}{$\checkmark$} & \textcolor{darkgreen}{$\checkmark$} & 23 \\
    B     &  \textcolor{red}{$\times$} & \textcolor{darkgreen}{$\checkmark$} & 22 \\
    C     & \textcolor{darkgreen}{$\checkmark$} & \textcolor{red}{$\times$} & 27 \\
    D     & \textcolor{red}{$\times$} & \textcolor{red}{$\times$} & 24 \\
    \bottomrule
    \end{tabular}
    \caption{Website versions configurations and response counts. The number of participants that tested each website version is shown in the last column. }
    \label{tab:website_version}
    \Description{Configuration and participant response counts of 4 versions of our user inference.}
\end{table}

We developed a pre-study questionnaire to gather insights into participants' prior experience with online clothing shopping which are detailed in the Appendix. Subsequently, we outlined a series of tasks for participants to undertake on our website, which has four versions as detailed in Table~\ref{tab:website_version}. Participants were assigned to test two versions each, of which the balanced assignment process is detailed in the Appendix. If the website version has size control, participants were asked to enter user true size for tops and bottoms and then perform the following:
\begin{enumerate}[itemsep=0pt, parsep=0pt]
\item Pick a top with your true size 
\item Try the top (with your true size) on 
\item Change the garment size of the top 
\item Try the top (with the changed size) on again 
\item Pick a bottom with your true size 
\item Try the bottom (with your true size) on 
\item Change the garment size of the bottom 
\item Try the bottom (with your true size) on again 
\item Continue from the result in step 6 (3rd result from the top) 
\item Try on the top from step 1 (with your true size) again 
\item Continue from the result in step 8 (3rd result from the top) 
\item Try on the top from step 3 (with a changed size) again
\end{enumerate}

Otherwise, they were asked to perform these tasks below:
\begin{enumerate}[itemsep=0pt, parsep=0pt]
\item Pick a top
\item Try the top on
\item Pick a bottom
\item Try the bottom on
\item Continue from the last result
\item Try the top from step 1 on again
\end{enumerate}
Similarity, if the website version used a self-image, participants were asked to update the photo taken at the start of the study, else they were asked to select an image from a set of model images displayed on the first page.

Following each website version's evaluation, participants were asked to complete a post-task questionnaire which comprised of three components: the NASA Task Load Index (TLX) questionnaire~\cite{hart2006nasa} with adjusted 11 gradations on the scales, a standard system usability questionnaire (SUS) in its positive version~\cite{lewis2018system}, and a questionnaire adapted from previous research~\cite{liu2020comparing} to examine how the various factors under investigation influenced user decision-making and overall experience. The questionnaire items were scored on a five-point Likert scale ranging from the option of ``1=Strongly disagree'' to ``5=Strongly agree'':
\begin{enumerate}[itemsep=0pt, parsep=0pt]
\item Shopping with this system was enjoyable for me.
\item I gain a sense of how the outfit might look on me.
\item This system helps me understand more about the appearance of the garments.
\item I feel confident that the clothes I choose are suitable for me.
\item This system would enhance the effectiveness of the shopping experience.
\item I want to use this system when I buy clothes online in the future.
\end{enumerate}

Our post-study questionnaire collect binary preference (website 1/website 2) from participants for enjoyment, better sense of fit, better understanding of garment appearance, confidence in choice, shopping effectiveness, and future usage. For the actual questions, please refer to the Appendix.
% Additionally, our post-study questionnaire included a pairwise comparison version of the aforementioned questions to directly contrast user experiences across the two website versions they tried. Each question offered a choice between ``the first website'' and ``the second website'':

% \begin{table}[]
% \centering
% \begin{tabular}{c|cc|cc}
% \toprule 
% \multirow{2}{*}{} & \multicolumn{2}{c|}{factor = size control} & \multicolumn{2}{c}{factor = upload self-image} \\ 
% \cmidrule(lr){2-3} \cmidrule(lr){4-5}
% & chi-square & p & chi-square & p \\  
% \midrule
% (1) & 4.1143  & \textbf{< 0.05} & 0.3462  & 0.5563 \\
% (2) & 7.3143  & \textbf{< 0.01} & 0.0385  & 0.8445 \\
% (3) & 13.8286 & \textbf{< 0.001} & 6.5000  & \textbf{< 0.05} \\
% (4) & 9.2571  & \textbf{< 0.01} & 4.6538  & \textbf{< 0.05} \\
% (5) & 2.8571  & 0.0910          & 3.1154  & 0.0776 \\
% (6) & 11.4286 & \textbf{< 0.001} & 5.7692  & \textbf{< 0.01} \\
% \bottomrule
% \end{tabular}

% \caption{Post-study response pair-wise comparison statistics. We calculate the McNemar chi-square value and p-value corresponding to each factor of interest with respect to each post-study question, where p-values that indicate statistical significance are highlighted in bold ($< 0.05$).}
% \label{tab:mcnemar}
% \Description{McNemar test of the post-study responses to show the significance of different user interface feature to each post-study question.}
% \end{table}

\begin{table}
\centering
\setlength{\tabcolsep}{3pt}
\begin{tabular}{c|ccc|ccc}
\toprule 
\multirow{2}{*}{} & \multicolumn{3}{c|}{factor = size control} & \multicolumn{3}{c}{factor = upload self-image} \\ 
\cmidrule(lr){2-4} \cmidrule(lr){5-7}
& chi-square & p & Cramer & chi-square & p & Cramer \\  \midrule
(1) & 5.050      & \textbf{$<$ 0.05} &  0.229          & 1.506      & 0.220    &   0.125         \\ 
(2) & 18.407     & \textbf{$<$ 0.001} &  \underline{0.438}          & 10.708     & \textbf{$<$ 0.01} &   \underline{0.334}         \\ 
(3) & 9.391      & \textbf{$<$ 0.01} &   \underline{0.313}         & 0.669      & 0.413    &   0.0834 \\ 
(4) & 12.062     & \textbf{$<$ 0.001} &  \underline{0.354}          & 8.199      & \textbf{$<$ 0.01} &    \underline{0.292} \\ 
(5) & 3.381      & 0.0660   &   0.187         & 6.024      & \textbf{$<$ 0.05} &     0.250       \\ 
(6) & 15.068     & \textbf{$<$ 0.001} &   \underline{0.396}         & 1.506      & 0.220    &     0.125 \\ 
\bottomrule
\end{tabular}

\caption{Post-study response statistics. We calculate the chi-square value and p-value corresponding to each factor of interest with respect to each post-study question, where p-values that indicate statistical significance are highlighted in bold ($< 0.05$), and Cramer's V values that indicate significant association are underlined ($> 0.25$) in the table.
}
\label{tab:chi_square}
\Description{Statistical analysis of the post-study responses to show the significance and association of different user interface feature to each post-study question.}
\end{table}

\begin{figure*}
    \centering
    \includegraphics[width=0.8\linewidth]{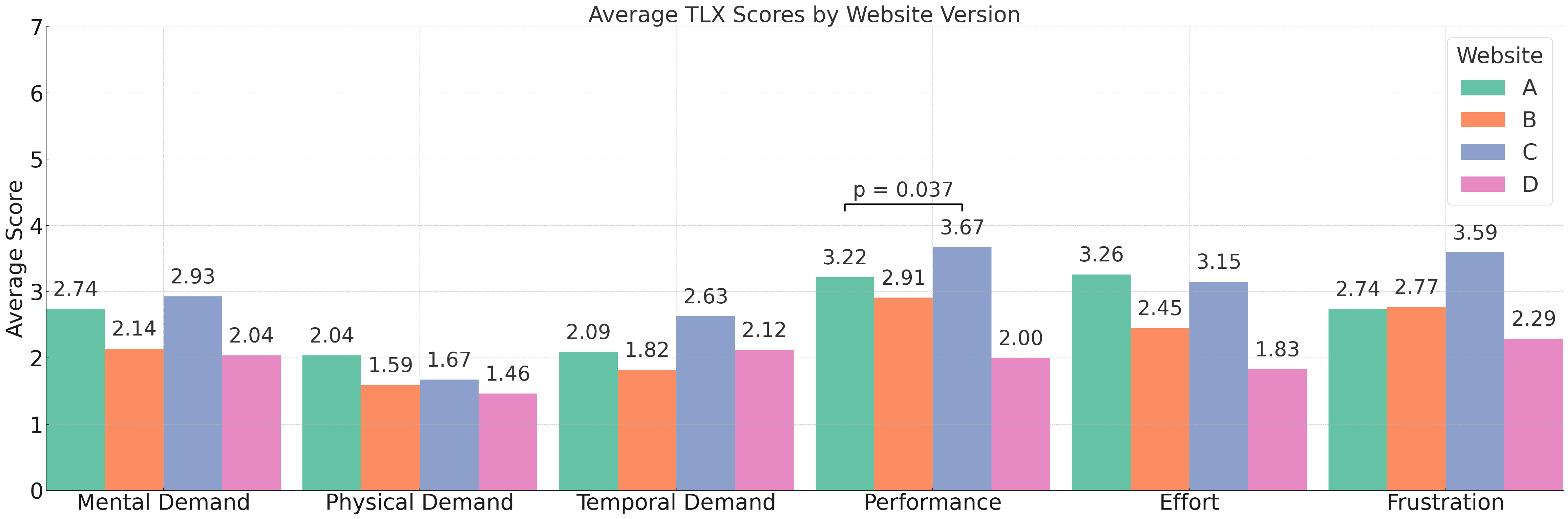}
    \caption{Task load index (TLX) metrics average across participant post-task responses. We calculated the average TLX metric scores across all participants, grouped by the website versions they tested. The results indicate that all website versions impose a nearly equal task load on users, with the exception of a significant increase in performance load observed when transitioning from website version A to version C ($p = 0.037$, see Appendix). This effect is particularly pronounced when self-images are not used.}
    \label{fig:results_website_version_tlx}
    \Description{Average task load index metrics across participant per user interface setting to show their effect on task load.}
\end{figure*}

\begin{figure*}
    \centering
    \includegraphics[width=\linewidth]{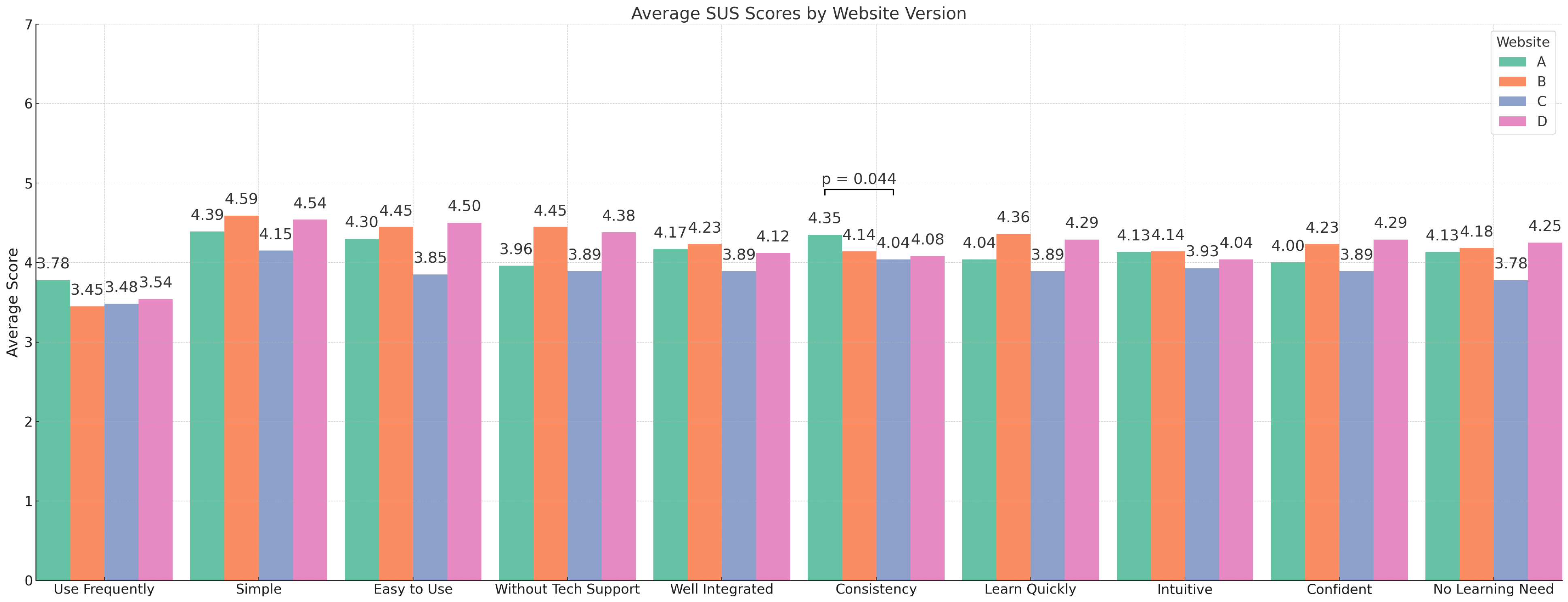}
    \caption{System usability scale (SUS) metrics average across participant post-task responses. We calculated the average SUS metric scores across all participants, grouped by the website versions they tested. The results indicate that all website versions demonstrate nearly equal usability, with the exception of a significant decrease in consistency observed when transitioning from website version A to version C ($p = 0.044$, see Appendix). This decrease is particularly notable when self-images are not utilized.}
    \label{fig:results_website_version_sus}
    \Description{Average system usability scale metrics across participant per user interface setting to show their effect on system usability.}
\end{figure*}

\subsection{Procedure}

Our 30-minute user study, approved by our local IRB (protocol \#26-24-0164), involved a single session per participant, who received \$10 USD as compensation. After participants provided informed consent, we took a full-body photograph of each participant. This photo was used if a task required image uploading. Participants evaluated two website versions through task sets, providing feedback via a questionnaire after each. A final questionnaire captured overall impressions of the study.

\section{Results}

\subsection{User Experience and Decision-making}

Table~\ref{tab:website_version} (Col.3) provides a breakdown of the frequency with which each website version was tested. Participant responses to the post-study questionnaire were analyzed, where chi-square values, p-values, and Cram\'er's V values are detailed in Table~\ref{tab:chi_square}. Statistically significant results ($\alpha < 0.05$) are highlighted in bold. The findings demonstrate that our size-controllable VTO significantly enhances user enjoyment, provides a clearer understanding of how chosen garments will appear on their bodies, aids in interpreting garment fit and appearance, and boosts confidence in garment selection. These factors collectively contribute to a strong preference for using our system for online clothing shopping.

To quantify the strength of associations between variables, we used Cram\'er's V, interpreting values around 0.3 as moderate associations and values around 0.5 as strong associations, with corresponding values underlined in the table. The analysis reveals a moderate to strong association between size-controllable VTO and an improved ability to visualize how clothing will look on users. Additionally, we observed moderate associations between size-controllable VTO and an enhanced understanding of garment appearance, as well as increased confidence in garment selection. Together, these factors contribute to users' preference for our website for future purchases.

The visualization of VTO results using participant self-images shows a significant positive impact. It indicates an improvement in user perception of how selected garments will look on them, a boost in their confidence in making garment selections, and enhancement of overall shopping experience. We identified a moderate association between the use of participant's own images and their ability to accurately gauge the appearance of outfits, as well as a reinforcement of confidence in the suitability of their selections.

Please refer to the Appendix for the corresponding contingency table and post-hoc analysis.

\subsection{Task-load Index}

To investigate the impact of size controllability and the use of participant self-images on cognitive load (measured using NASA TLX), we computed the average TLX scores for each website version (Fig.~\ref{fig:results_website_version_tlx}) and conducted an aligned rank transform mixed-design ANOVA analysis. The analysis revealed a significant negative impact ($p = 0.037$) on the TLX performance metric when transitioning from website version A to C. This suggests that enabling size control in VTO and/or using participant self-images to visualize VTO results did not increase task load.

For further details, please refer to Appendix.

\subsection{System Usability}

To investigate the impact of size controllability and the use of participant self-images on system usability (measured using SUS), we computed the average SUS scores for each website version (Fig.~\ref{fig:results_website_version_sus}). Responses were recorded on a Likert scale ranging from ``1 = Strongly disagree'' to ``5 = Strongly agree.''

An ANOVA analysis revealed that only the transition from website version A to version C, which used generic model images instead of self-images, showed a significant negative impact on the SUS consistency metric ($p = 0.044$). No other significant impacts were observed, indicating that enabling size control in VTO and/or using participant self-images to visualize VTO results did not negatively affect overall system usability.

For further details, please refer to Appendix.

\section{Limitations and Future Work}

While SiCo's size controllability feature indicates effectiveness in our user study, our participant pool was limited to individuals recruited from a university campus. As a result, the study under-represents certain age groups, occupations, financial backgrounds, and body types/sizes. Future work should aim to conduct a more comprehensive study with broader demographic coverage to better understand the real-world applicability of our system and assess its robustness across customers with different appearances.

In terms of user experience, our system significantly enhances user ability to visualize how outfits might appear on their bodies, increasing their confidence in selecting preferred sizes. By enabling users to make real-time size and fit adjustments, our approach bridges the critical gap between perceived fit in a virtual setting and the actual garment fit, potentially reducing costly returns. However, while this increased confidence is promising, its effect on actual purchase behavior, such as lowering return rates, has not yet been verified. Our current study relies on subjective participant judgments, which are valuable indicators of user experience, but not sufficient to assess long-term behavioral outcomes. To more rigorously evaluate real-world impact, future studies could incorporate real garments for in-person try-ons, allowing users to compare physical fit with VTO visualizations side by side. Additionally, integrating our VTO system into a live retail environment would enable measurement of longitudinal purchase behavior and return rates.

Our system currently does not allow users to style the same garment in different ways, such as unbuttoning a shirt or rolling up pant legs. Implementing this feature would require modifications to our garment mask, which we plan to incorporate in future iterations. Additionally, the system does not yet support layering more than two garments. While users can technically try on multiple top garments (e.g., a shirt and a jacket), without options to ``unbutton'' or ``unzip'' the jacket, the shirt underneath remains mostly obscured.

Beyond these limitations, it is important to acknowledge that our user identity preservation mechanism cannot fully eliminate all biases inherent in the generative model, Stable Diffusion. For instance, biases related to skin color may persist if individuals with different skin tones are underrepresented in the training data. This could result in slight alterations to user skin tones, potentially causing offense. Addressing this issue will require re-training or fine-tuning the model using more diverse and inclusive datasets.

Finally, while our use of a diffusion-based generative model achieves higher quality and greater generality, it introduces a drawback in terms of speed. Users currently experience an average wait time of approximately one minute for each VTO result, with the system running on an NVIDIA A6000 in a lab setting. In addition to the latency, the hardware demands and associated electricity consumption raise concerns about scalability and environmental impact. However, we argue that this could be mitigated in future work by adopting faster diffusion-based models, such as distilled models, on cloud services. Nevertheless, assessing the system’s net environmental impact requires in-field testing to account for real-world usage patterns and energy offsets.

\section{Conclusion}

In this paper, we introduced a web-based virtual try-on (VTO) system featuring a size-controllable VTO backbone, designed to provide users with a familiar online clothing shopping experience enhanced by our innovative VTO functionality. To assess the impact of size controllability and the use of self or model images for clothing visualization on user experience and decision-making, we conducted a user study ($n = 48$). The results revealed a strong preference for our VTO feature with size adjustment. Participants reported that this functionality significantly improved their ability to visualize clothing on themselves, enhanced their perception of garment appearance, and increased their confidence in making informed purchase decisions.

\bibliographystyle{ACM-Reference-Format}
\bibliography{sample-base}

\appendix

\appendix
% \section*{Appendix}

\section{Pre-Study Questionnaire Analysis\label{sec:pre_study_questionnaire_analysis}}

\begin{figure}
\centering
\setlength{\tabcolsep}{1pt}
\begin{tabular}{cc}
 \includegraphics[width=0.4\linewidth]{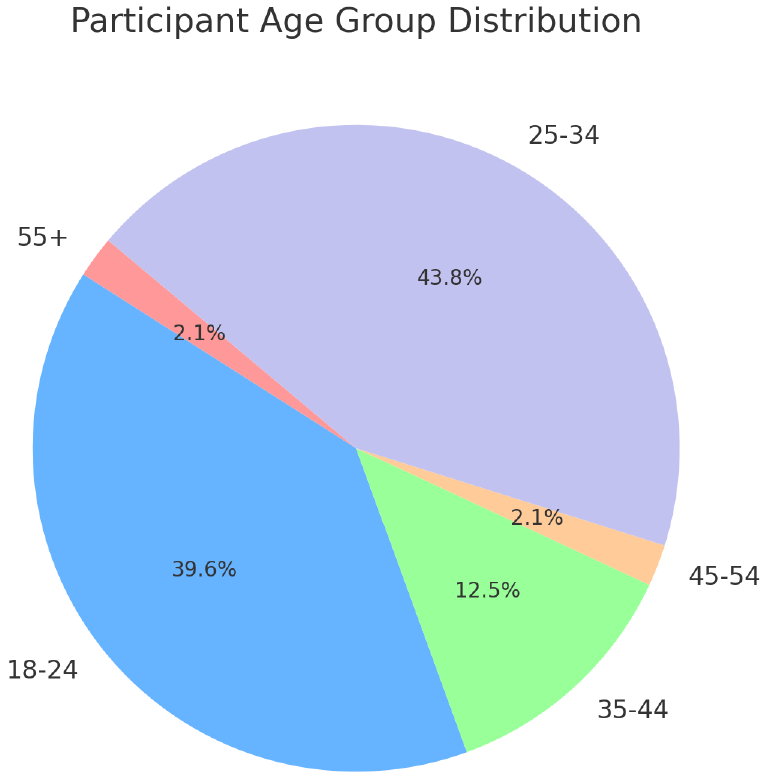}  &   \includegraphics[width=0.45\linewidth]{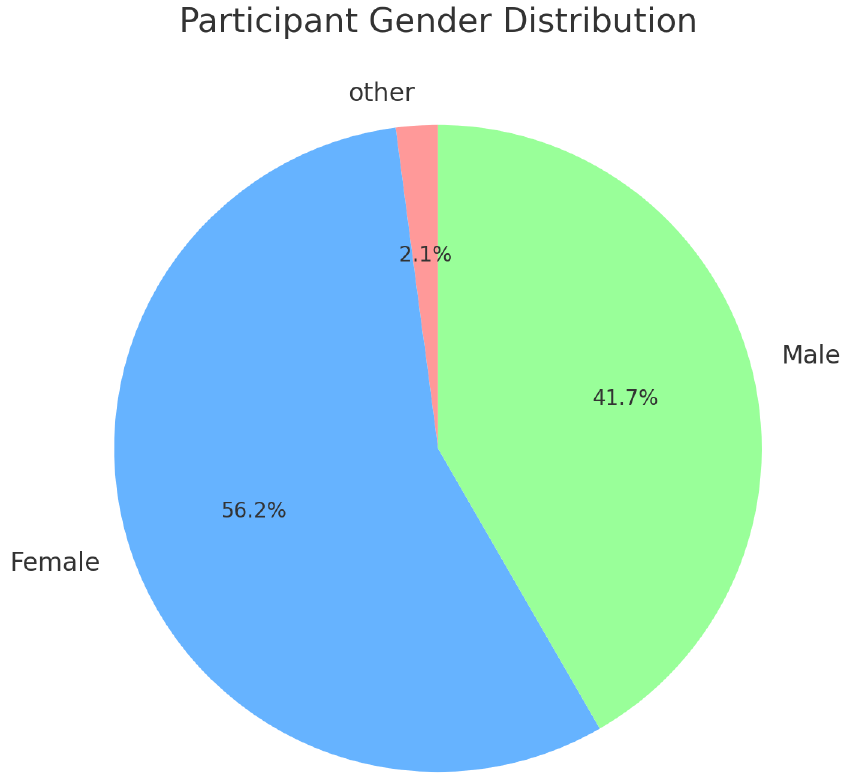} \\ 
 \includegraphics[width=0.5\linewidth]{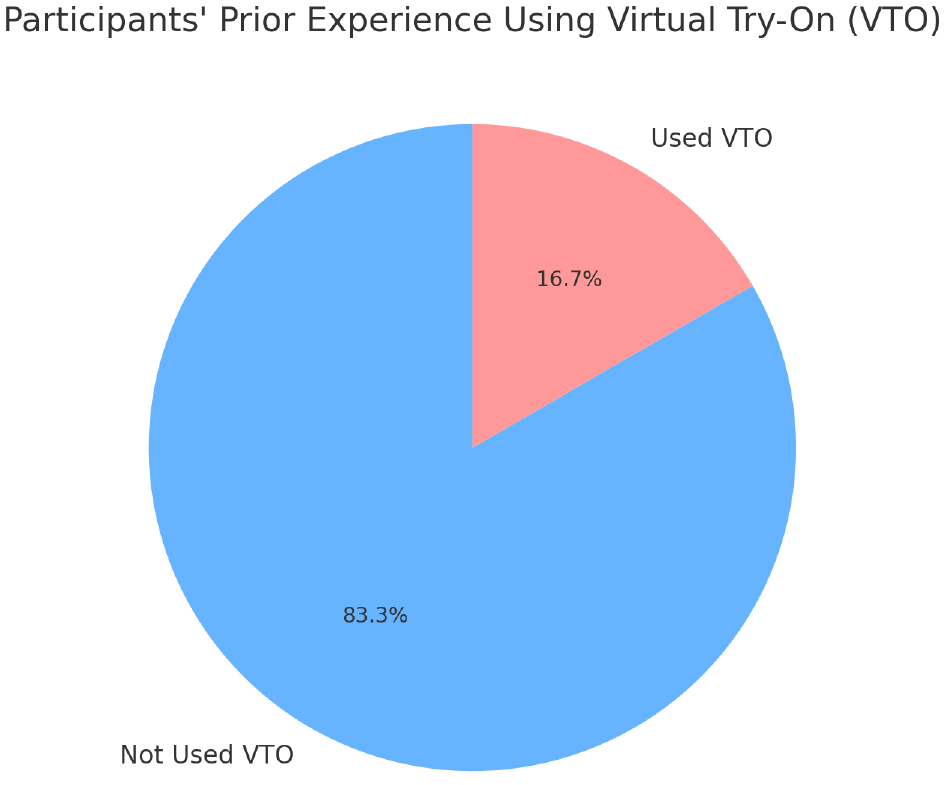} &
  \includegraphics[width=0.45\linewidth]{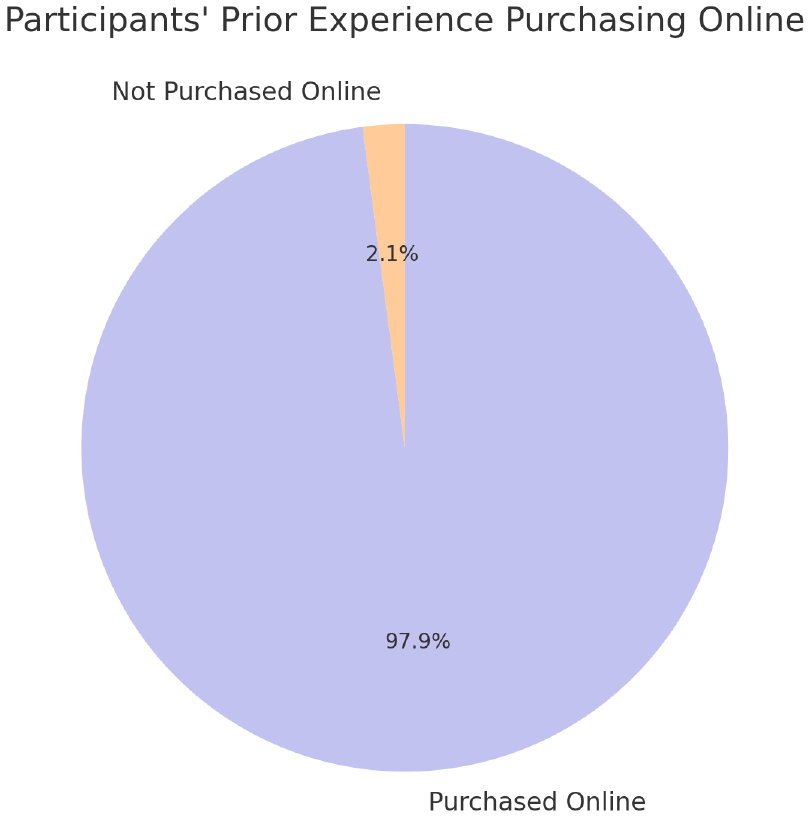} \\
  \includegraphics[width=0.49\linewidth]{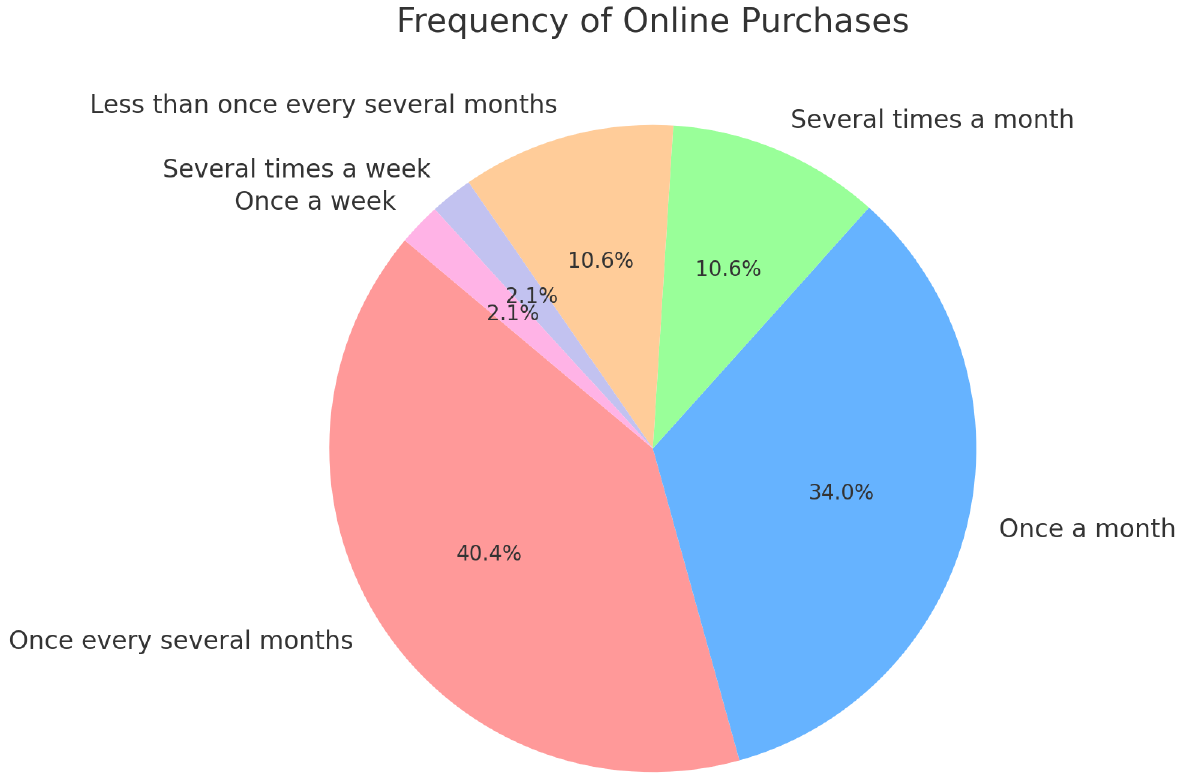} & 
  \includegraphics[width=0.49\linewidth]{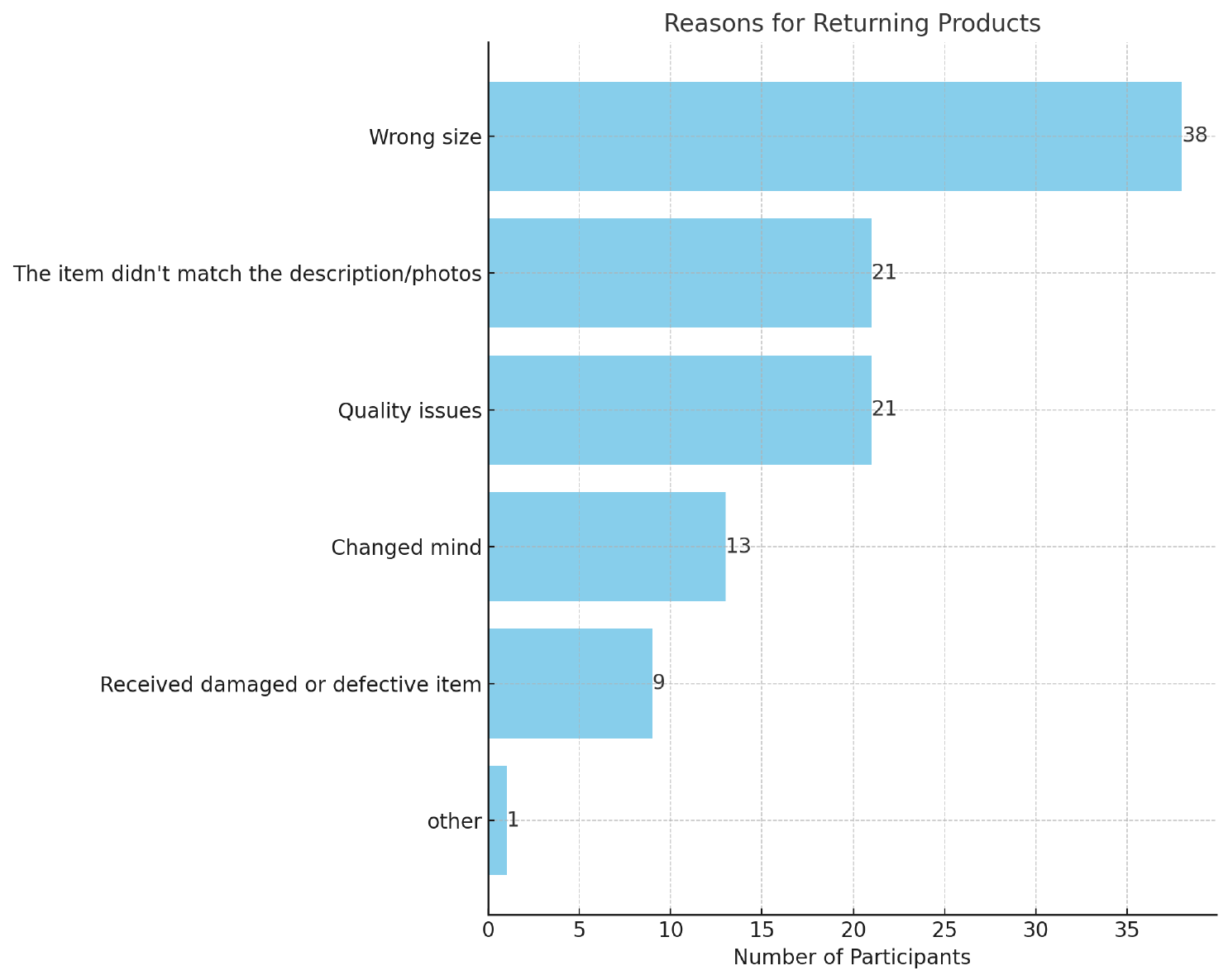} \\ 
\end{tabular}
\caption{Participant pre-study questionnaire response statistics. We provide participant age group, gender, prior experience with online clothing shopping and returns, as well as using virtual try-on service.}
\label{fig:pre-study}
\Description{participant pre-study questionnaire statistics for 6 questions. For participant age group distribution, 43.8\% are from age group 25-34, 39.6\% are from 18-24. For participant gender, 56.2\% are female and 41.7\% are male. 83.3\% of the participants haven't used a virtual try-on service before. 97.9\% of the participants have purchased clothing online before. For online shopping frequency, 40.4\% shop once every several months and 34.0\% shop once per month. ``Wrong sizes'' is the leading reason for returns, where 38 of all participants select this.}
\end{figure}

We recruited 48 participants to test SiCo, with Figure~\ref{fig:pre-study} showing a breakdown of responses from the pre-study questionnaire. Most participants reported prior experience with online clothing shopping for themselves, with the most common shopping frequency being ``once every several months'' (43.1\%). A smaller subset of participants (15.4\%) had prior experience using a VTO service. When asked about reasons for returning items purchased online, the most common response was ``wrong size,'' with 39 out of 48 participants indicating this as a factor.

Each participant was assigned two website versions to test in sequential order based on their arrival to the study. Assignments followed a cyclic pattern: the first participant tested the versions listed under index 1 in Table~\ref{tab:website_index}, the second participant tested the versions under index 2, and so on. This cycle repeated until the 25th participant, who returned to test the versions under index 1. This process ensured a uniform distribution of evaluations across all website versions.

\begin{table}[b]
    \centering
    \begin{tabular}{c|cc||c|cc||c|cc}
    \toprule
    index & $1^{st}$ & $2^{nd}$ & index & $1^{st}$ & $2^{nd}$ & index & $1^{st}$ & $2^{nd}$ \\ \midrule
    1 & D & A & 9  & B & C & 17 & C & D \\  
    2 & A & D & 10 & B & A & 18 & C & B \\
    3 & D & A & 11 & B & C & 19 & C & D \\  
    4 & A & D & 12 & B & A & 20 & C & B \\
    5 & A & C & 13 & C & A & 21 & D & C \\
    6 & A & B & 14 & B & D & 22 & D & B \\
    7 & A & C & 15 & C & A & 23 & D & C \\
    8 & A & B & 16 & B & D & 24 & D & B \\
    \bottomrule
    \end{tabular}
    \caption{Participant website versions assignment. Participants were assigned to test two versions each, in sequential order based on their arrival to the study. The assignment process was cyclic, starting with the first participant evaluating the versions listed under index 1, the second participant evaluating the versions under index 2, and so on. This pattern was repeated until the 25th participant, who circled back to test the versions under index 1, ensuring a uniform distribution of evaluations across all website versions.}
    \label{tab:website_index}
    \Description{Our strategy to assign participant version of two user inferences they will be testing to ensure uniform distribution across user inference settings we are interested in analyzing.}
\end{table}

\section{Post-Study Questionnaire Analysis}

Our post-study questionnaire asks participants for their preferences among two websites they tested:
\begin{enumerate}[itemsep=0pt, parsep=0pt]
\item \_\_\_\_\_ is more enjoyable to use.
\item \_\_\_\_\_ gives me a better sense of how the outfit might look on me.
\item \_\_\_\_\_ helps me understand more about the appearance of the garments better.
\item I am confident that the clothes I choose suit me with \_\_\_\_\_.
\item \_\_\_\_\_ would enhance the effectiveness of the shopping experience more.
\item I prefer to use \_\_\_\_\_ when I buy clothes online in the future.
\end{enumerate}

The $2 \times 2$ contingency tables comparing the use of a size-controllable Virtual Try-On (VTO) versus a baseline VTO as well as the use of the participant’s image versus a selected model image are presented in Table~\ref{tab:contingency} We perform chi-square post-hoc analysis with adjusted residuals (Table~\ref{tab:post_hoc}) where we observe that more participants than expected (under the null hypothesis of no association) preferred the
version with size-controllability VTO and using their own images for virtual try-on, respectively.

\begin{table*}
\centering
\small
\begin{tabular}{c|c|cc|c|cc}
\toprule 
question  & \multicolumn{3}{c|}{factor = size control}    & \multicolumn{3}{c}{factor = upload self-image}            \\ \midrule
\multirow{3}{*}{(1) enjoyable} & factor   & preferred & not preferred & factor   & preferred & not preferred \\  \cmidrule(lr){2-4} \cmidrule(lr){5-7}
& included & 31       & 19            & included &  26   & 19            \\  
& excluded & 17        & 29            & excluded & 22   & 29 \\ \midrule

\multirow{3}{*}{(2) sense-look} & factor   & preferred & not preferred & factor   & preferred & not preferred \\  \cmidrule(lr){2-4} \cmidrule(lr){5-7}
& included & 36        & 14            & included &  31   & 14            \\  
& excluded & 12       & 34            & excluded & 17   & 34 \\ \midrule

\multirow{3}{*}{(3) appearance} & factor   & preferred & not preferred & factor   & preferred & not preferred \\  \cmidrule(lr){2-4} \cmidrule(lr){5-7}
& included & 33        & 17            & included &  25   & 20            \\  
& excluded & 15        & 31           & excluded & 23   & 28 \\ \midrule

\multirow{3}{*}{(4) suitability} & factor   & preferred & not preferred & factor   & preferred & not preferred \\  \cmidrule(lr){2-4} \cmidrule(lr){5-7}
& included & 34       & 16            & included &  30   & 15            \\  
& excluded & 14        & 32           & excluded & 18  & 33 \\ \midrule

\multirow{3}{*}{(5) effectiveness} & factor   & preferred & not preferred & factor   & preferred & not preferred \\  \cmidrule(lr){2-4} \cmidrule(lr){5-7}
& included & 30        & 20            & included &  29   & 16            \\  
& excluded & 18        & 28           & excluded & 19   & 32 \\ \midrule

\multirow{3}{*}{(6) future-use} & factor   & preferred & not preferred & factor   & preferred & not preferred \\  \cmidrule(lr){2-4} \cmidrule(lr){5-7}
& included & 35        & 15            & included &  26   & 19            \\  
& excluded & 13        & 33           & excluded & 22   & 29 \\ \bottomrule
\end{tabular}

\caption{Post-study response contingency table. We generate the contingency tables that count the number of responses for each question in our post-study questionnaire, grouped by the factor of interest (with/without size control, uploading user self-image vs. using a selected model image).}
\label{tab:contingency}
\Description{Contingency table of participant post-study responses grouped by user interface settings to show the effect of each setting to user experience.}
\end{table*}

\begin{table*}
\centering
\small
\begin{tabular}{c|ccc|ccc}
\toprule
question & \multicolumn{3}{c|}{factor = size control}                                     & \multicolumn{3}{c}{factor = upload self-image}            \\ \midrule
\multirow{3}{*}{(1) enjoyable} & factor   & preferred & not preferred & factor   & preferred & not preferred \\   \cmidrule(lr){2-4} \cmidrule(lr){5-7}
& included & 1.200       & -1.200            & included &  0.738   & -0.738            \\  
& excluded & -1.251        & 1.251           & excluded & -0.693   & 0.693 \\ \midrule

\multirow{3}{*}{(2) sense-look} & factor   & preferred & not preferred & factor   & preferred & not preferred \\   \cmidrule(lr){2-4} \cmidrule(lr){5-7}
& included & 2.200        & -2.200            & included &  1.791   & -1.791            \\  
& excluded & -2.394       & 2.394            & excluded & -1.683   & 1.683 \\ \midrule

\multirow{3}{*}{(3) appearance} & factor   & preferred & not preferred & factor   & preferred & not preferred \\   \cmidrule(lr){2-4} \cmidrule(lr){5-7}
& included & 1.600        & -1.600            & included &  1.668   & -1.668            \\  
& excluded & -0.527       & 0.527           & excluded & -0.495   & 0.495 \\ \midrule

\multirow{3}{*}{(4) suitability} & factor   & preferred & not preferred & factor   & preferred & not preferred \\   \cmidrule(lr){2-4} \cmidrule(lr){5-7}
& included & 1.800       & -1.800            & included &  1.581   & -1.581            \\  
& excluded & -1.877        & 1.877           & excluded & -1.485  & 1.485 \\ \midrule

\multirow{3}{*}{(5) effectiveness} & factor   & preferred & not preferred & factor   & preferred & not preferred \\   \cmidrule(lr){2-4} \cmidrule(lr){5-7}
& included & 1.000        & -1.000           & included & 1.370   &   -1.370          \\  
& excluded &  -1.043        & 1.043           & excluded & -1.287   & 1.287 \\ \midrule

\multirow{3}{*}{(6) future-use} & factor   & preferred & not preferred & factor   & preferred & not preferred \\   \cmidrule(lr){2-4} \cmidrule(lr){5-7}
& included & 2        & -2            & included &  0.738   & -0.738            \\  
& excluded & -2.085        & 2.085           & excluded & -0.693   & 0.693 \\ \bottomrule
\end{tabular}
\caption{Post-study response post-hoc analysis. We compute the adjusted residual table where we observe that more participants than expected (under the null hypothesis of no association) preferred the version with size-controllability VTO and using their own images for virtual try-on, respectively. }
\label{tab:post_hoc}
\Description{Post-hoc analysis of participant post-study responses grouped by user interface settings to show the effect of each setting to user experience.}
\end{table*}

\begin{figure}
    \centering
    \begin{tabular}{@{}c@{}c@{}c@{}}
    \includegraphics[width=0.49\linewidth]{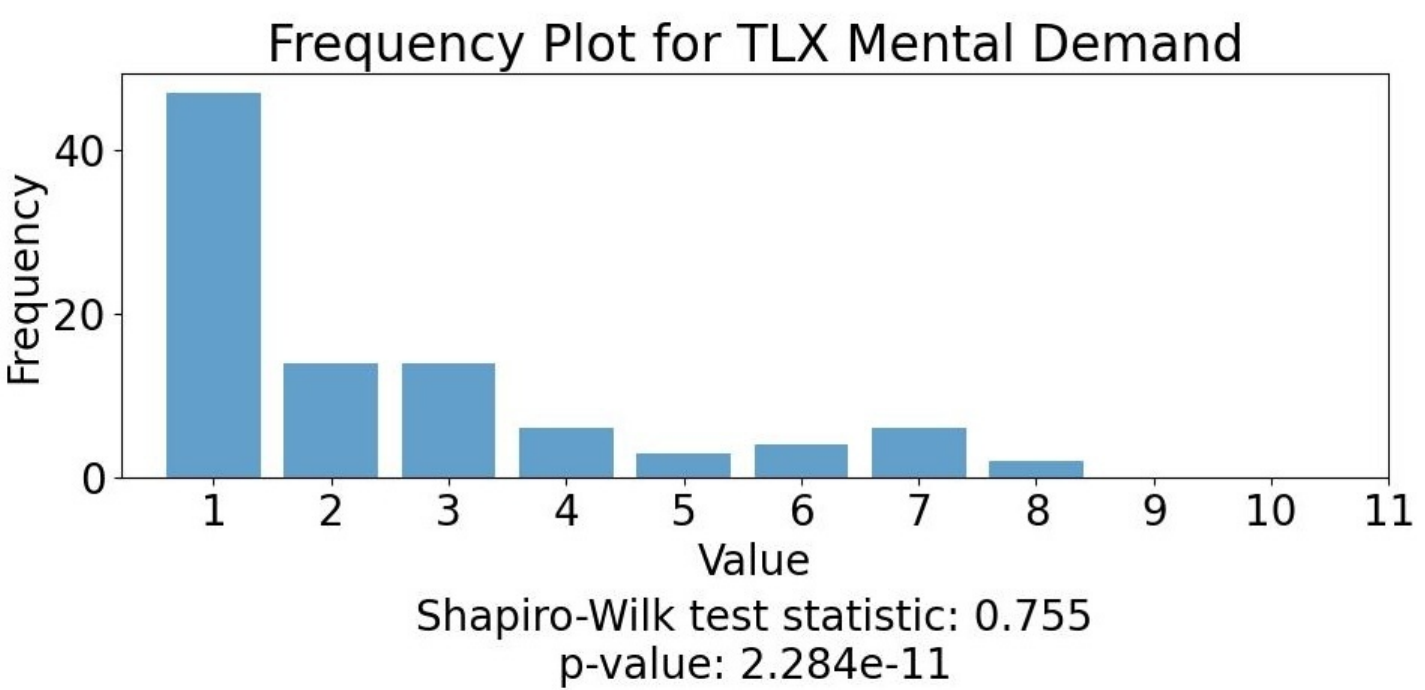} & \includegraphics[width=0.49\linewidth]{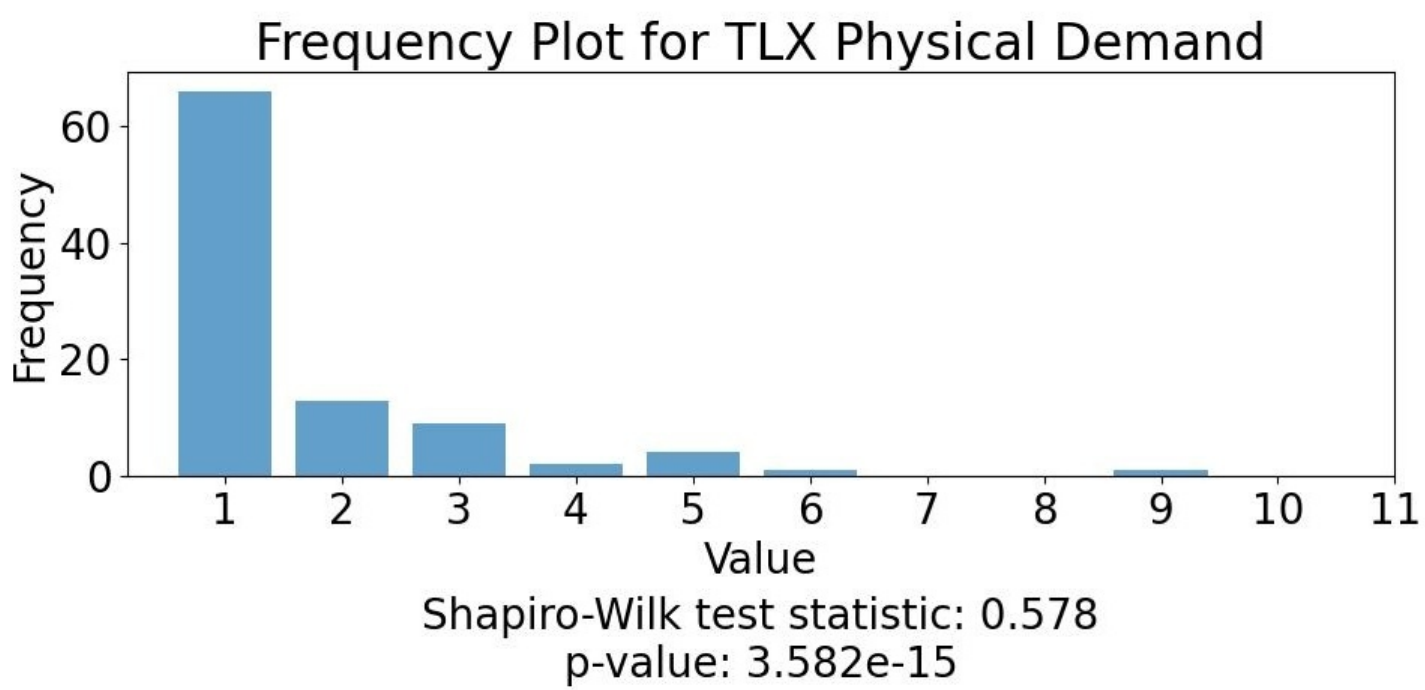} \\
    \includegraphics[width=0.49\linewidth]{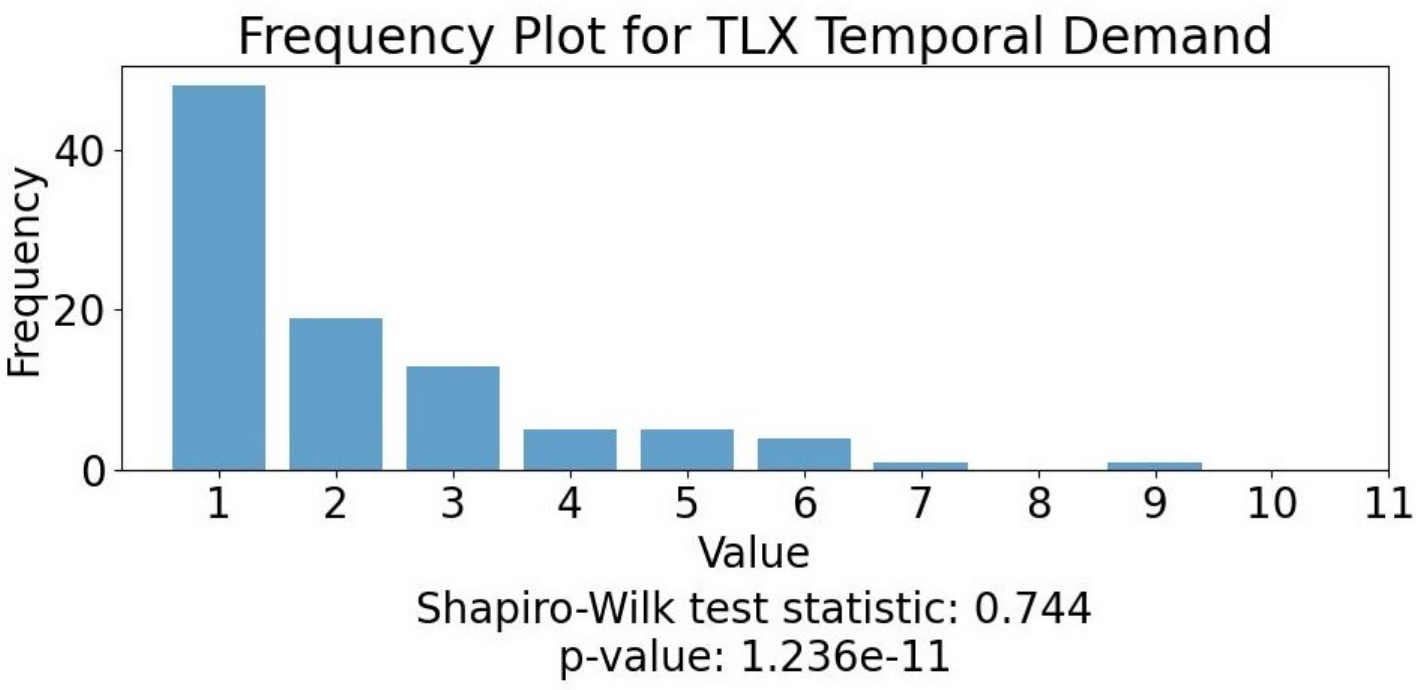} &
     \includegraphics[width=0.49\linewidth]{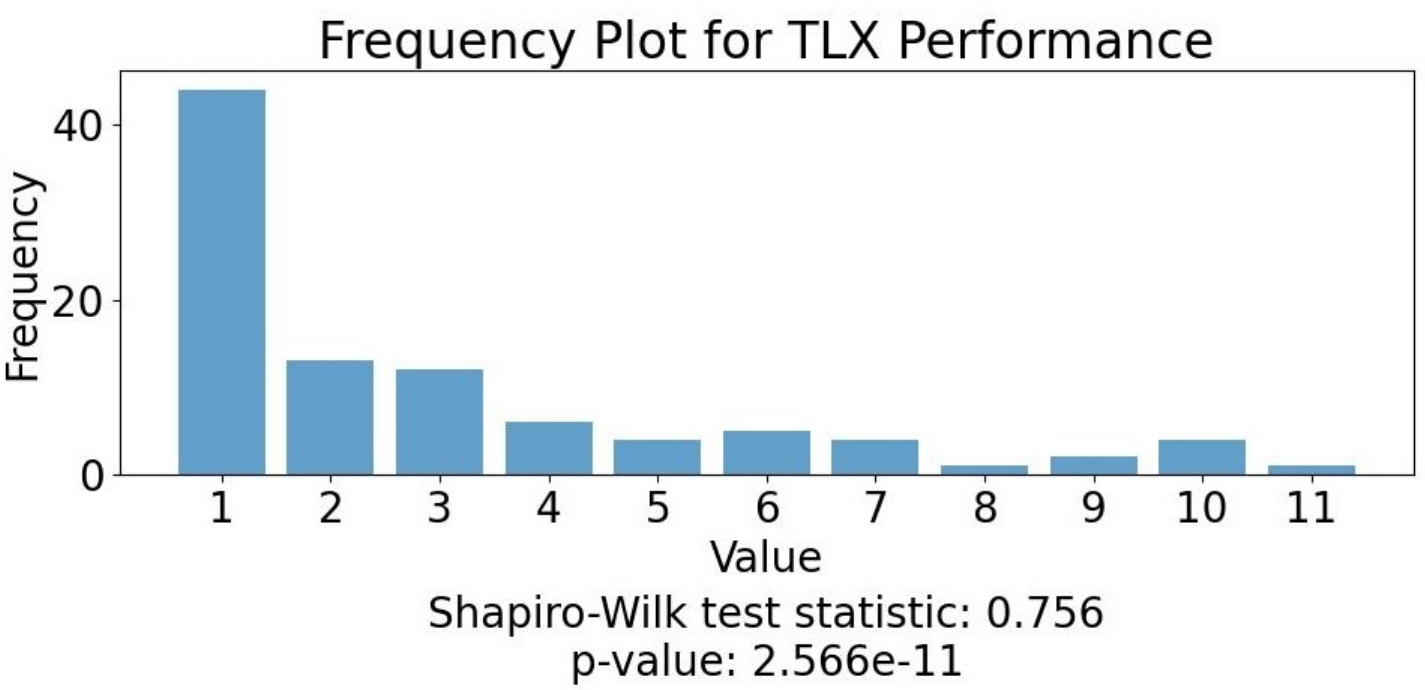}  \\
     \includegraphics[width=0.49\linewidth]{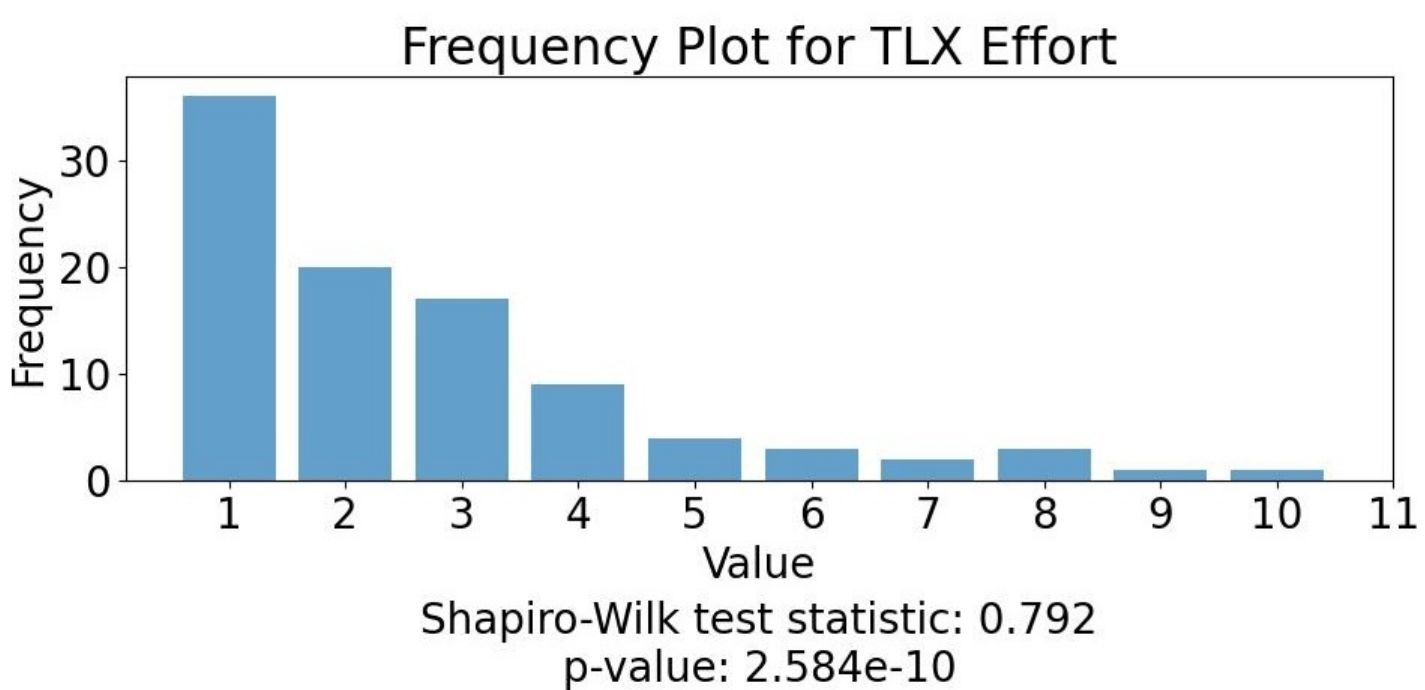} & \includegraphics[width=0.49\linewidth]{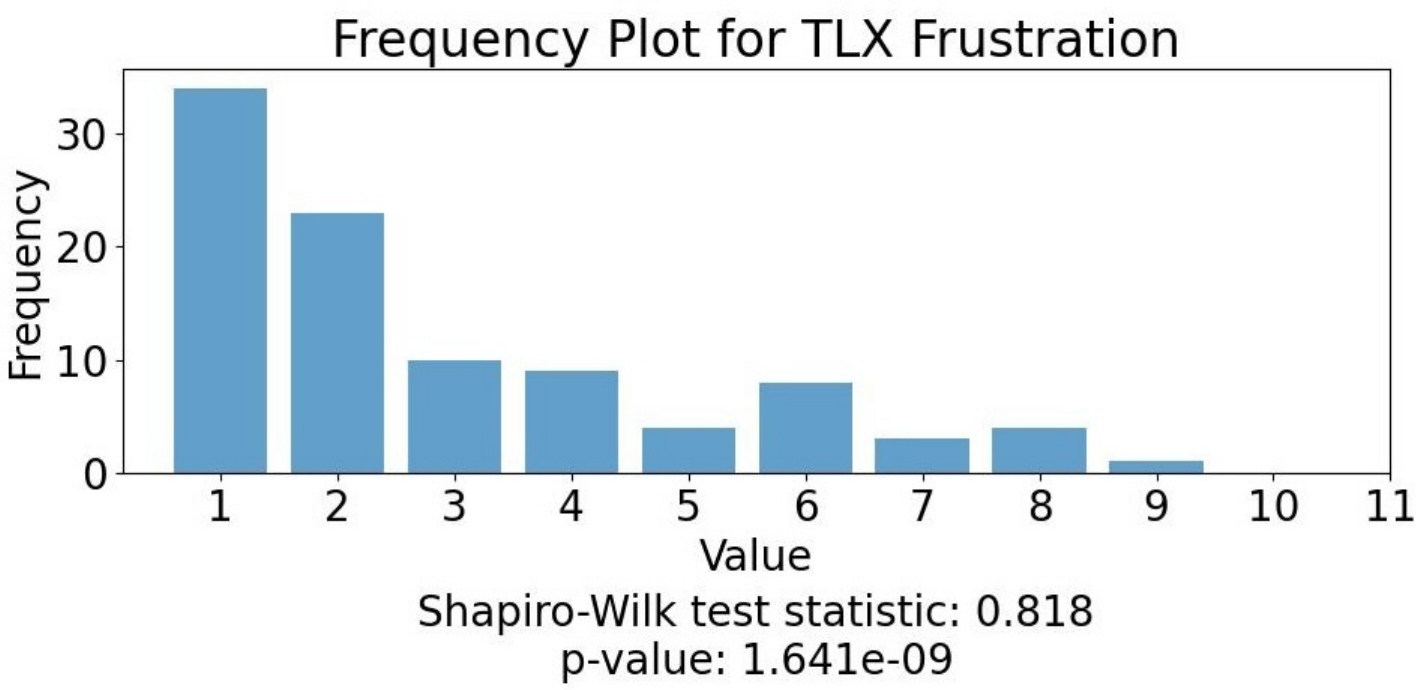} \\
    \end{tabular}
    \caption{Post-task task load index (TLX) response. Here we show the frequency plots of TLX responses of all participants across four website versions.}
    \label{fig:tlx}
    \Description{Frequency bar plot for post-task task load index response. For each metric, 1 (low task load) is the most frequently selected answer.}
\end{figure}

\begin{figure}
\centering
\begin{tabular}{cc}
\includegraphics[width=0.49\linewidth]{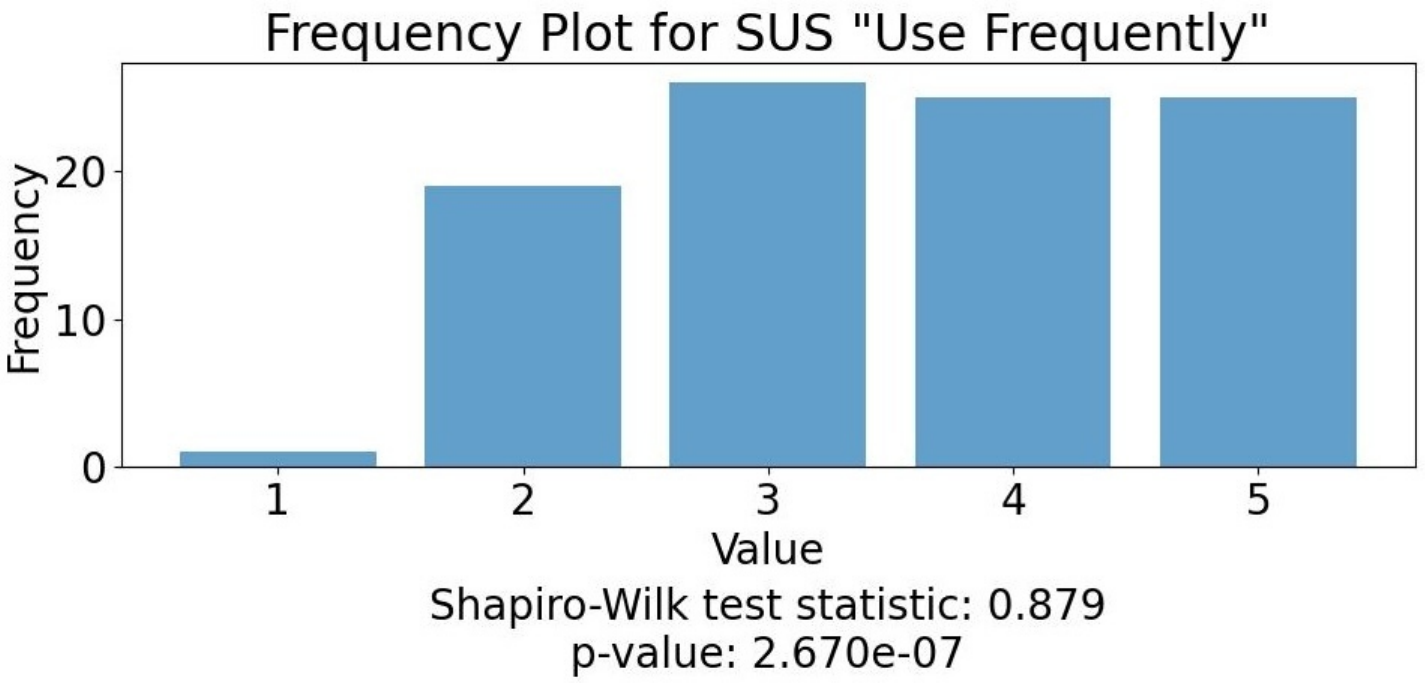} & 
\includegraphics[width=0.49\linewidth]{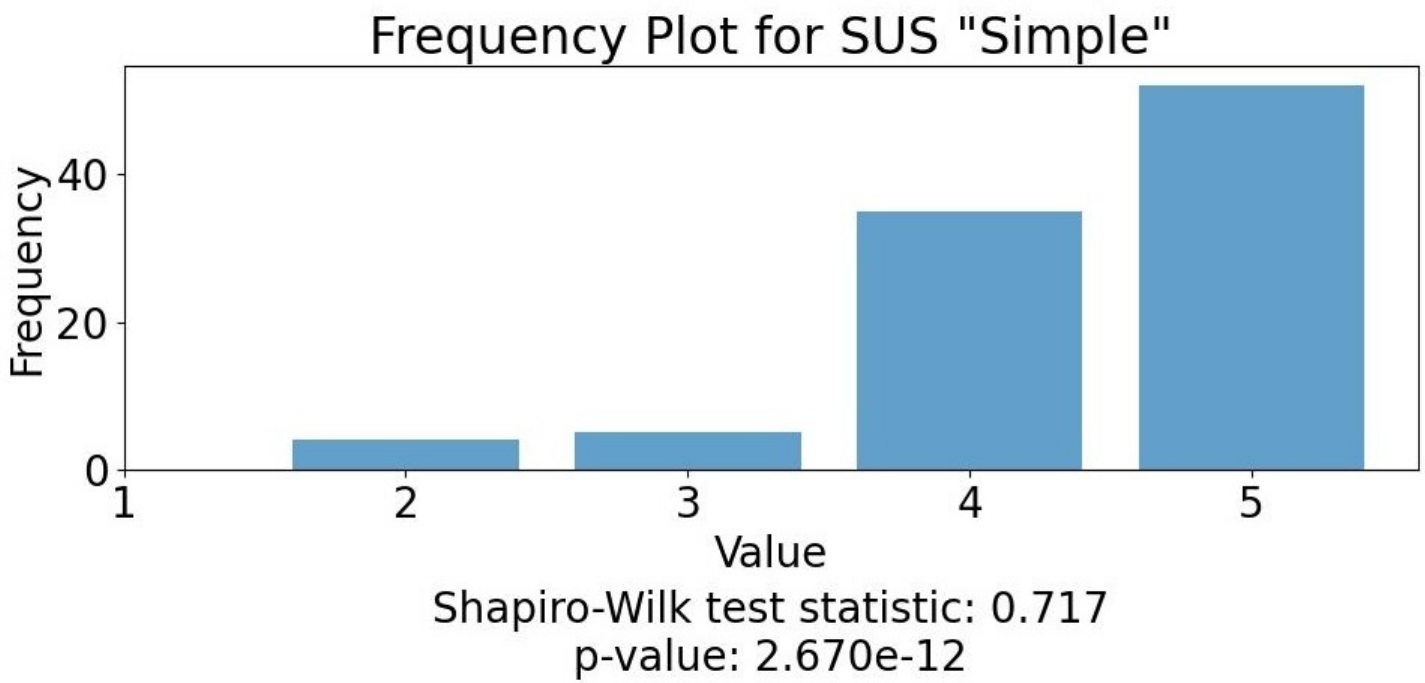} \\
\includegraphics[width=0.49\linewidth]{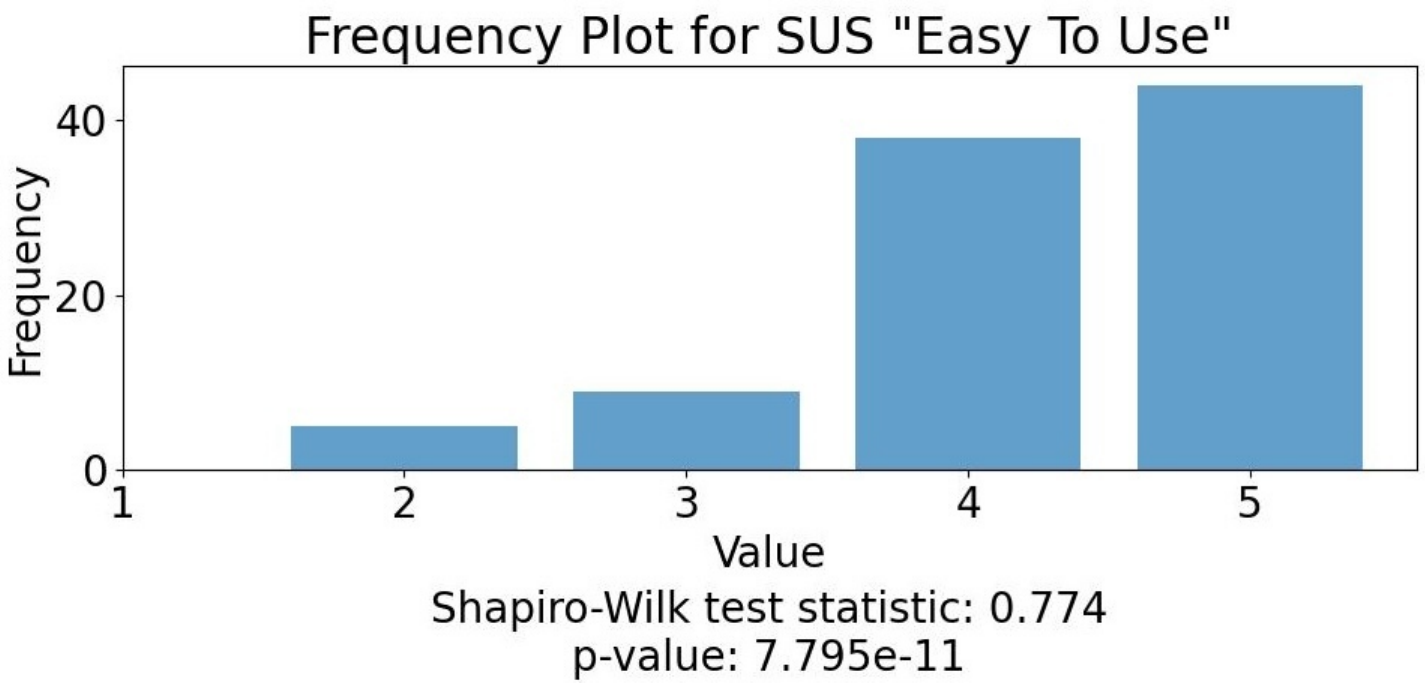} & 
\includegraphics[width=0.49\linewidth]{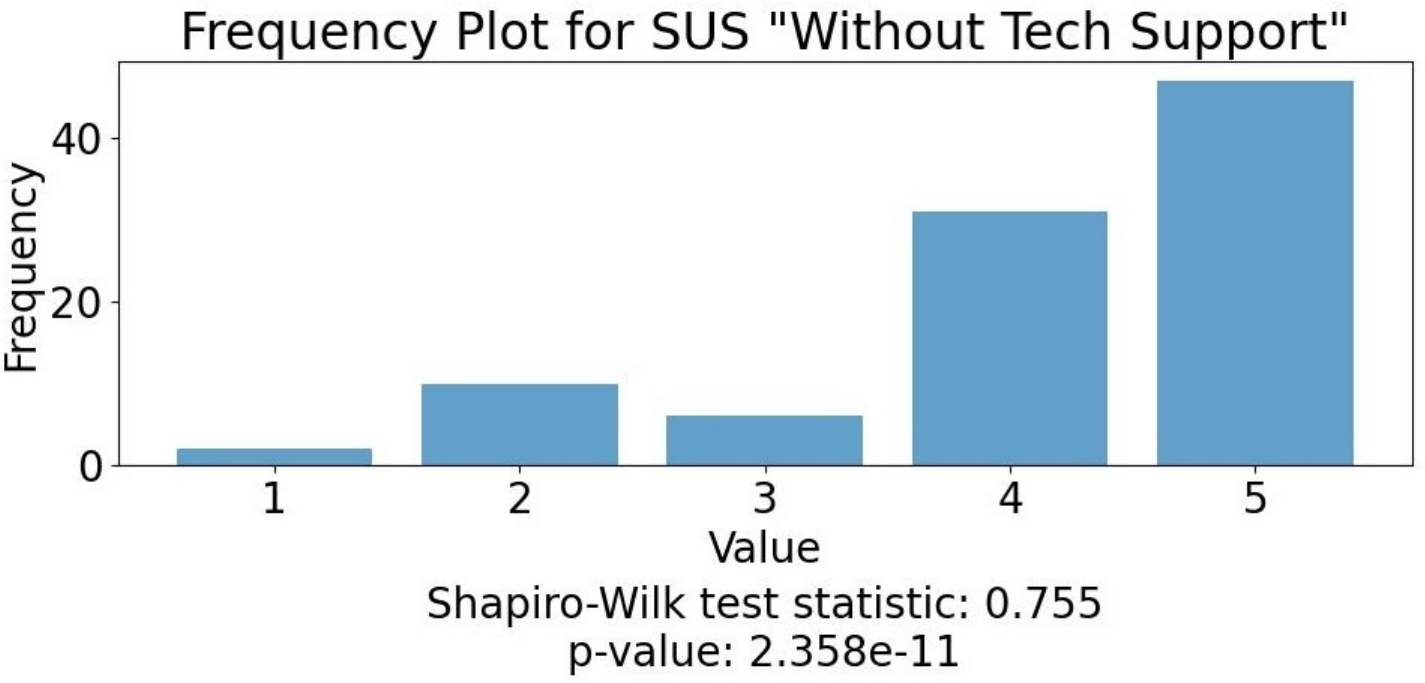} \\
\includegraphics[width=0.49\linewidth]{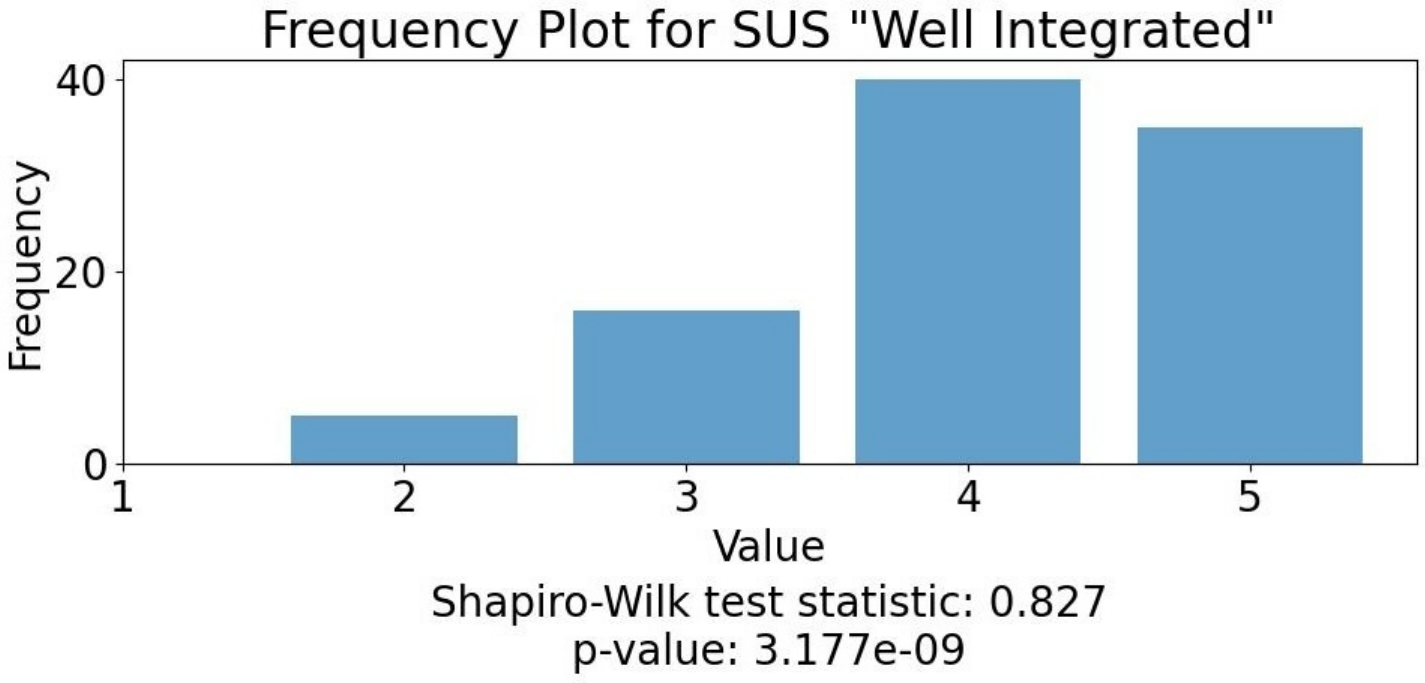} & 
\includegraphics[width=0.49\linewidth]{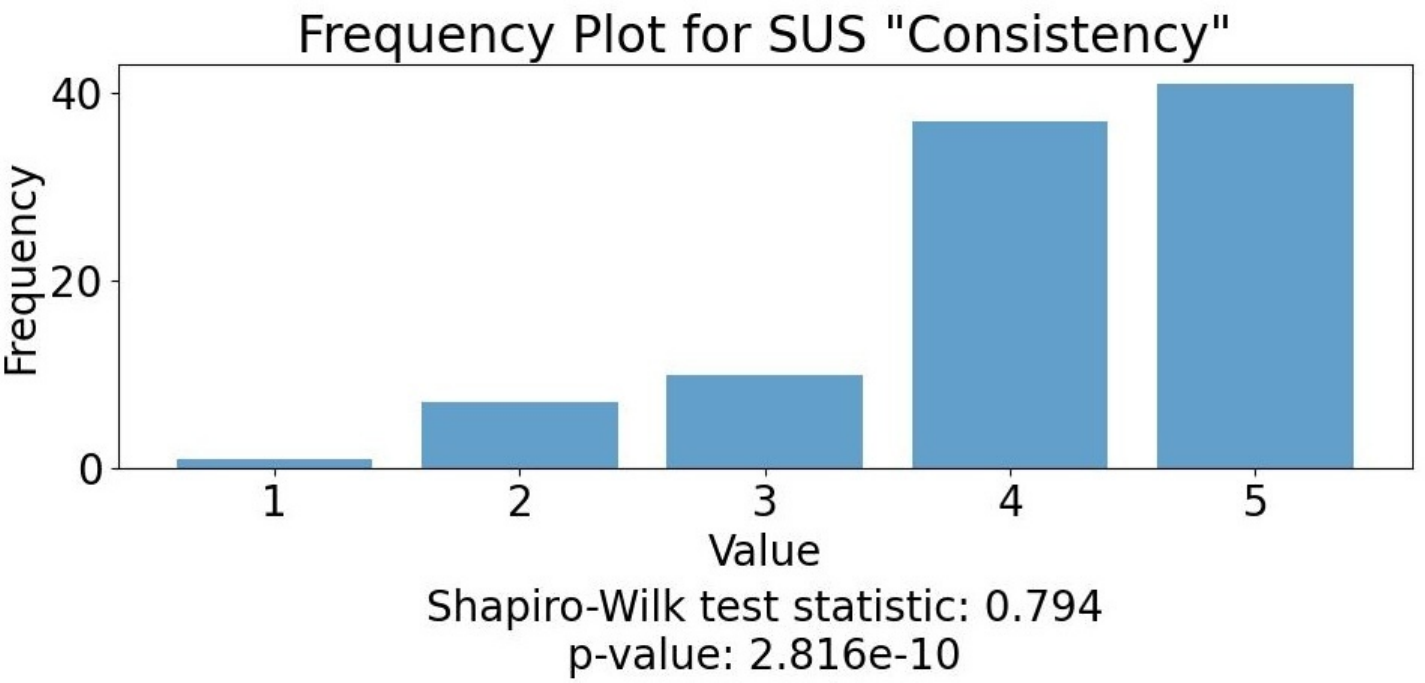} \\
\includegraphics[width=0.49\linewidth]{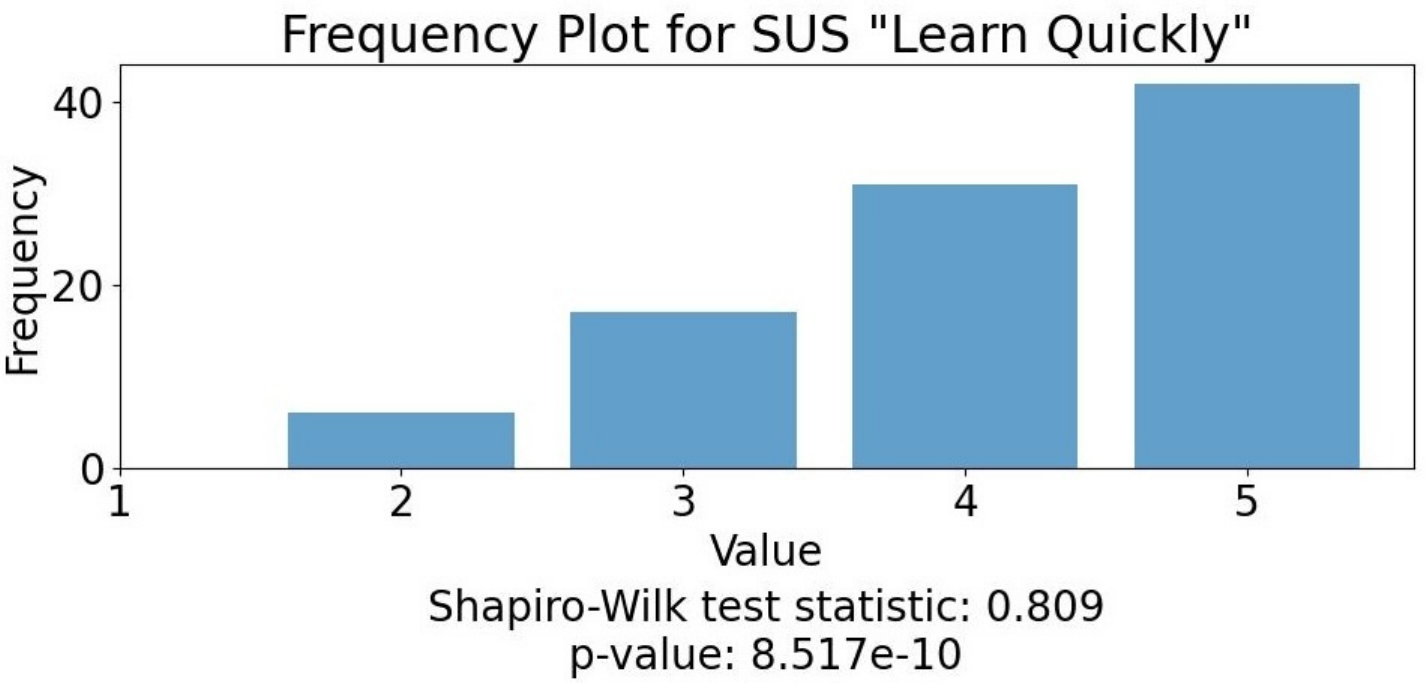} & 
\includegraphics[width=0.49\linewidth]{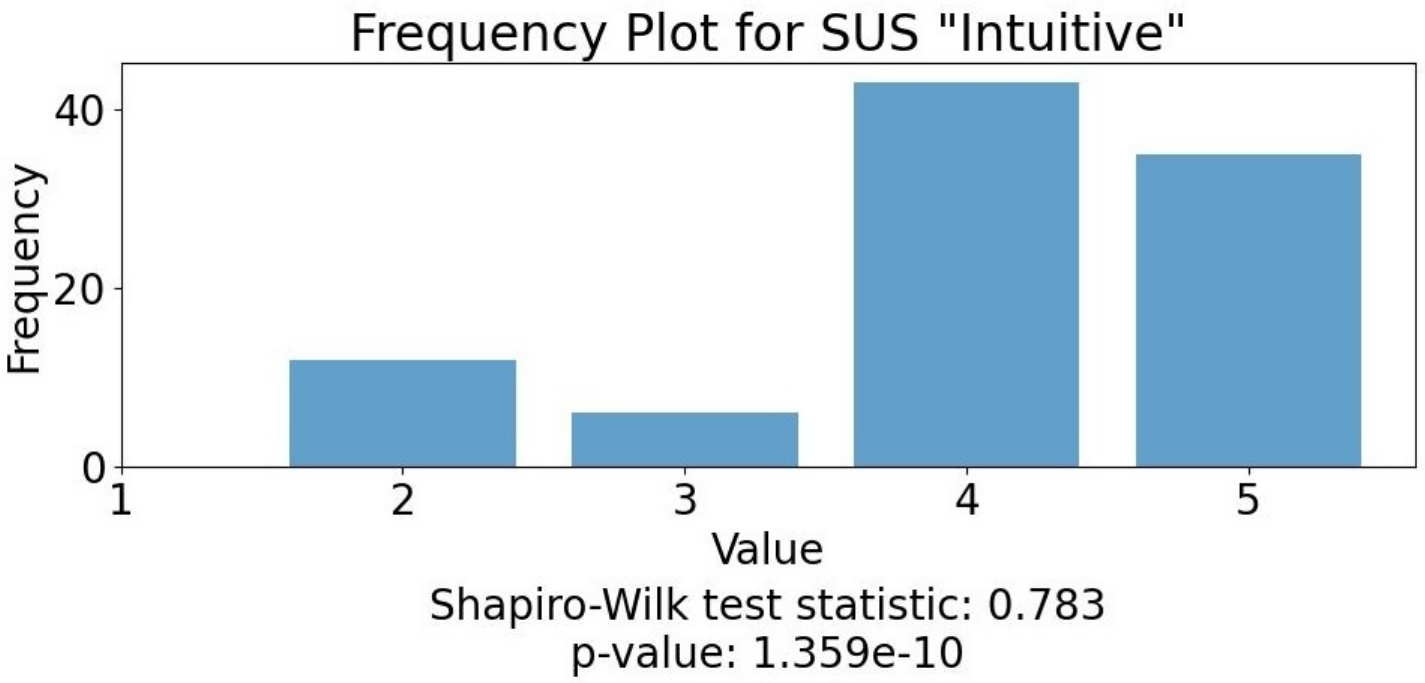} \\
\includegraphics[width=0.49\linewidth]{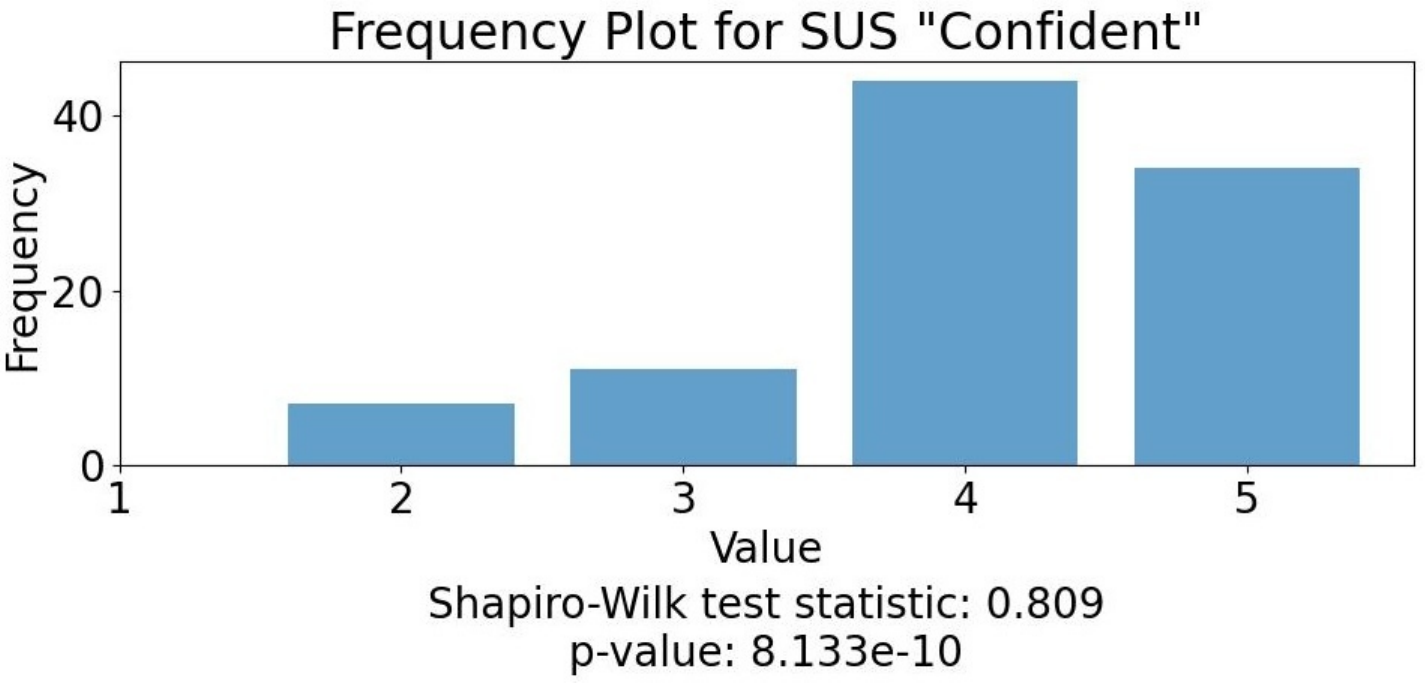} & 
\includegraphics[width=0.49\linewidth]{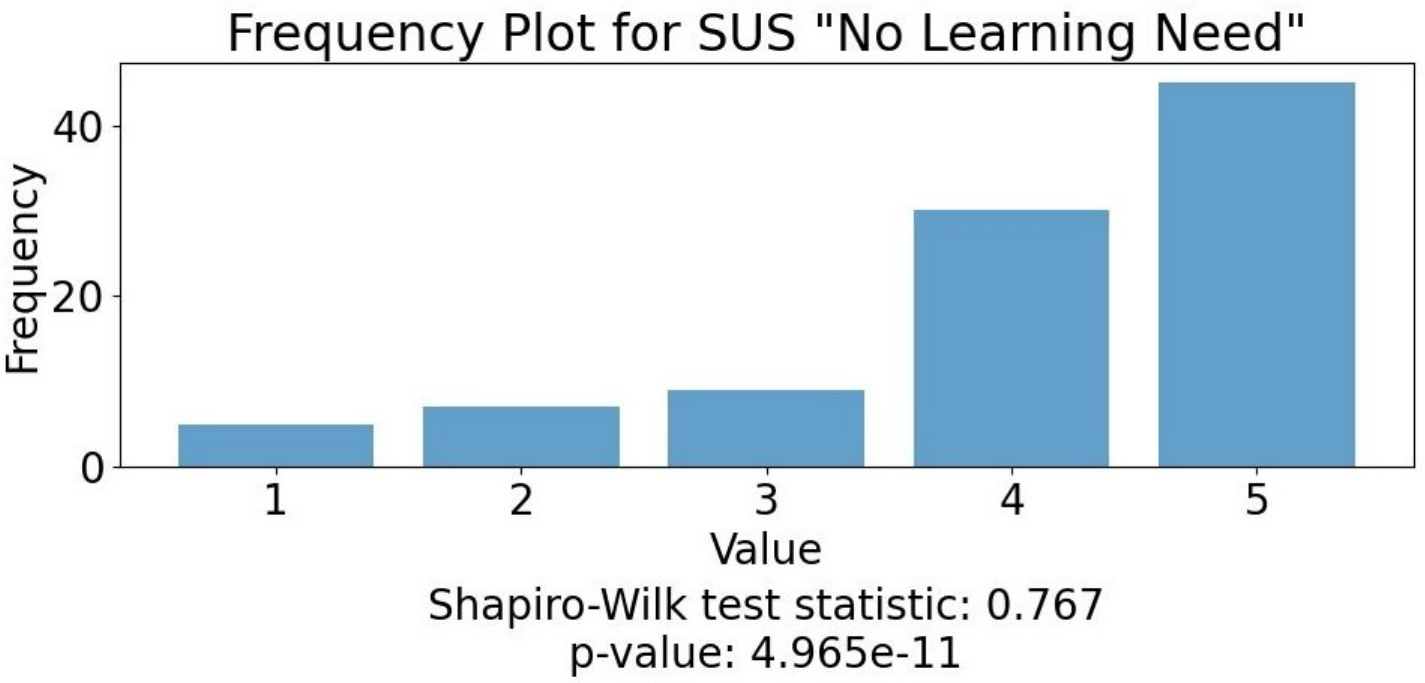} \\
\end{tabular}
\caption{Post-task system usability scale (SUS) response. Frequency plots of SUS responses across four website versions.}
\label{fig:sus}
\Description{Frequency bar plot for post-task system usability scale response. For most metric, 5 (high usability) is the most frequent answer.}
\end{figure}

\begin{table*}
\centering
\begin{tabular}{c|ccccccc}
\toprule
 &  & Coef. & Std.Err. & z & P$>|$z$|$ & [0.025 & 0.975] \\ \cmidrule(lr){2-8}
 & Intercept & 12.406 & 1.119 & 11.088 & 0.000 & 10.213 & 14.599 \\
 & condition[T.B] & -1.622 & 1.039 & -1.561 & 0.119 & -3.659 & 0.415 \\
 & condition[T.C] & 1.624 & 1.018 & 1.595 & 0.111 & -0.371 & 3.620 \\
 \multirow{-5}{*}{\centering Mental Demand} & condition[T.D] & 0.327 & 1.068 & 0.306 & 0.759 & -1.766 & 2.420 \\

\midrule
 &  & Coef. & Std.Err. & z & P$>|$z$|$ & [0.025 & 0.975] \\ \cmidrule(lr){2-8}
 & Intercept & 12.124 & 1.098 & 11.037 & 0.000 & 9.971 & 14.277 \\
 & condition[T.B] & -0.521 & 1.326 & -0.393 & 0.694 & -3.120 & 2.078 \\
 & condition[T.C] & 1.686 & 1.289 & 1.308 & 0.191 & -0.840 & 4.212 \\
 \multirow{-5}{*}{\centering Physical Demand} & condition[T.D] & 0.378 & 1.343 & 0.281 & 0.778 & -2.255 & 3.011 \\

\midrule
 &  & Coef. & Std.Err. & z & P$>|$z$|$ & [0.025 & 0.975] \\ \cmidrule(lr){2-8}
 & Intercept & 11.905 & 1.125 & 10.584 & 0.000 & 9.701 & 14.110 \\
 & condition[T.B] & 1.033 & 1.061 & 0.973 & 0.330 & -1.047 & 3.112 \\
 & condition[T.C] & 1.026 & 1.039 & 0.987 & 0.323 & -1.011 & 3.063 \\
 \multirow{-5}{*}{\centering Temporal Demand}  & condition[T.D] & 0.570 & 1.088 & 0.524 & 0.601 & -1.563 & 2.702 \\

\midrule
 &  & Coef. & Std.Err. & z & P$>|$z$|$ & [0.025 & 0.975] \\ \cmidrule(lr){2-8}
 & Intercept & 11.222 & 1.216 & 9.232 & 0.000 & 8.840 & 13.605 \\
 & condition[T.B] & 0.497 & 1.330 & 0.374 & 0.708 & -2.109 & 3.104 \\
 & condition[T.C] & 2.710 & 1.298 & 2.089 & \textbf{0.037} & 0.167 & 5.253 \\
 \multirow{-5}{*}{\centering Performance} & condition[T.D] & 1.898 & 1.363 & 1.393 & 0.164 & -0.772 & 4.569 \\

\midrule
 &  & Coef. & Std.Err. & z & P$>|$z$|$ & [0.025 & 0.975] \\ \cmidrule(lr){2-8}
 & Intercept & 12.185 & 1.248 & 9.761 & 0.000 & 9.738 & 14.632 \\ 
 & condition[T.B] & -0.939 & 1.428 & -0.657 & 0.511 & -3.737 & 1.860 \\
 & condition[T.C] & 1.344 & 1.392 & 0.965 & 0.334 & -1.385 & 4.073 \\
\multirow{-5}{*}{\centering Effort} & condition[T.D] & 0.900 & 1.454 & 0.619 & 0.536 & -1.949 & 3.748 \\
\midrule

 &  & Coef. & Std.Err. & z & P$>|$z$|$ & [0.025 & 0.975] \\ \cmidrule(lr){2-8}
 & Intercept & 12.248 & 1.308 & 9.363 & 0.000 & 9.684 & 14.812 \\
 & condition[T.B] & -0.436 & 1.573 & -0.277 & 0.782 & -3.518 & 2.646 \\
 & condition[T.C] & 0.988 & 1.537 & 0.642 & 0.521 & -2.025 & 4.000 \\
\multirow{-5}{*}{\centering Frustration} & condition[T.D] & 0.589 & 1.594 & 0.370 & 0.712 & -2.534 & 3.712 \\
\bottomrule
\end{tabular}
\caption{Aligned rank transform mixed-design ANOVA for task load index (TLX). We calculate the aligned rank transform mixed-design ANOVA for each TLX metric, where the values with statistical significance ($\alpha = 0.05$) are highlighted in bold.}
\label{tab:art_anova_tlx}
\Description{Statistical analysis of participant post-study responses via ANOVA grouped by user interface settings to show the effect of each setting to task load.}
\end{table*}

\section{TLX Statistical Analysis}

We want to investigate the impact of size-controllability and the use of participant self-images on task load using NASA TLX. Given that the reported scores do not adhere to a normal distribution (as indicated by the Shapiro-Wilk test and p-values shown below each plot in the Fig.~\ref{fig:tlx}), we conducted an aligned rank transform mixed-design ANOVA analysis as summarized in Tab.~\ref{tab:art_anova_tlx}, where we present the baseline values (Intercept) for website version A and the estimated value changes with website versions B-D. 

Looking at the z and p-values, used for indicating significance, only the change from website version A to version C showed a significant negative impact ($p = 0.037$) on the TLX performance metric. No other significant impacts were observed, suggesting that enabling size-control in VTO and/or using participant self-images to visualize VTO results did not affect task load.

\section{SUS Statistical Analysis}

To investigate the impact of size-controllability and the use of participant self-images on system usability (measured using SUS), we conduct an aligned rank transform mixed-design ANOVA analysis. The results of this analysis are summarized in Table~\ref{tab:art_anova_sus} given that the reported scores do not adhere to a normal distribution (as indicated by the Shapiro-Wilk test and p-values shown below each plot in Fig.~\ref{fig:sus}). Here we present the baseline values (Intercept) for website version A and the estimated changes with versions B-D. 

Upon examination of the z and p-values, used for indicating significance, we find that only the change from website version A to version C showed a significant negative impact ($p = 0.044$) on the SUS consistency metric. No other significant impacts were observed, suggesting that enabling size-control in VTO and/or using participant self-images to visualize VTO results did not affect system usability.

\begin{table*}
\centering
\small
\begin{tabular}{c|c|c|cccccc}
\toprule
& & & Coef. & Std.Err. & z & P$>|$z$|$ & [0.025 & 0.975] \\ \cmidrule(lr){3-9}
& & Intercept & 12.637 & 1.257 & 10.057 & 0.000 & 10.174 & 15.100 \\
& & condition[T.B] & -1.220 & 1.408 & -0.867 & 0.386 & -3.981 & 1.540 \\
& & condition[T.C] & 1.714 & 1.372 & 1.249 & 0.212 & -0.975 & 4.403 \\
\multirow{-5}{*}{\centering Use Frequently} & \multirow{-5}{*}{\parbox{3cm}{\centering I think that I would like to use the website frequently.}} & condition[T.D] & -1.065 & 1.445 & -0.737 & 0.461 & -3.898 & 1.767 \\

\midrule
& & & Coef. & Std.Err. & z & P$>|$z$|$ & [0.025 & 0.975] \\ \cmidrule(lr){3-9}
& & Intercept & 12.071 & 1.171 & 10.310 & 0.000 & 9.776 & 14.365 \\
& & condition[T.B] & -0.240 & 1.363 & -0.176 & 0.860 & -2.912 & 2.432 \\
& & condition[T.C] & 1.945 & 1.327 & 1.466 & 0.143 & -0.655 & 4.546 \\
\multirow{-5}{*}{\centering Simple} & \multirow{-5}{*}{\parbox{3cm}{\centering I found the website to be simple.}} & condition[T.D] & 0.041 & 1.387 & 0.030 & 0.976 & -2.677 & 2.760 \\

\midrule
& & & Coef. & Std.Err. & z & P$>|$z$|$ & [0.025 & 0.975] \\ \cmidrule(lr){3-9}
& & Intercept & 12.288 & 1.310 & 9.383 & 0.000 & 9.721 & 14.855 \\
& & condition[T.B] & -1.078 & 1.716 & -0.628 & 0.530 & -4.442 & 2.286 \\
& & condition[T.C] & 1.959 & 1.645 & 1.191 & 0.234 & -1.266 & 5.184 \\
\multirow{-5}{*}{\centering Easy To Use} & \multirow{-5}{*}{\parbox{3cm}{\centering I thought the website was easy to use.}} & condition[T.D] & -0.078 & 1.719 & -0.045 & 0.964 & -3.448 & 3.292 \\

\midrule
 &  & Coef. & Std.Err. & z & P$>|$z$|$ & [0.025 & 0.975] \\ \cmidrule(lr){3-9}
 & & Intercept & 13.079 & 1.196 & 10.935 & 0.000 & 10.735 & 15.424 \\
 & & condition[T.B] & -2.330 & 1.309 & -1.779 & 0.075 & -4.896 & 0.237 \\
&  & condition[T.C] & 1.976 & 1.267 & 1.561 & 0.119 & -0.506 & 4.459 \\
\multirow{-5}{*}{Without Tech Support} & \multirow{-5}{*}{\parbox{3cm}{\centering I think that I could use the website without the support of a technical person.}} & condition[T.D] & -2.114 & 1.351 & -1.565 & 0.118 & -4.760 & 0.533 \\

\midrule
& & & Coef. & Std.Err. & z & P$>|$z$|$ & [0.025 & 0.975] \\ \cmidrule(lr){3-9}
& & Intercept & 12.618 & 1.222 & 10.327 & 0.000 & 10.224 & 15.013 \\
& & condition[T.B] & -1.240 & 1.374 & -0.902 & 0.367 & -3.934 & 1.453 \\
& & condition[T.C] & 1.743 & 1.338 & 1.303 & 0.193 & -0.880 & 4.367 \\
\multirow{-5}{*}{\centering Well Integrated} & \multirow{-5}{*}{\parbox{3cm}{\centering I found the various functions in the website were well integrated.}} & condition[T.D] & -1.007 & 1.410 & -0.714 & 0.475 & -3.770 & 1.756 \\

\midrule
& & & Coef. & Std.Err. & z & P$>|$z$|$ & [0.025 & 0.975] \\ \cmidrule(lr){3-9}
& & Intercept & 11.832 & 1.237 & 9.562 & 0.000 & 9.407 & 14.257 \\
& & condition[T.B] & -0.632 & 1.481 & -0.427 & 0.670 & -3.535 & 2.271 \\
& & condition[T.C] & 2.921 & 1.447 & 2.019 & \textbf{0.044} & 0.085 & 5.757 \\
\multirow{-5}{*}{\centering Consistency}  & \multirow{-5}{*}{\parbox{3cm}{\centering I thought there was a lot of consistency in the website.}}  & condition[T.D] & 0.256 & 1.502 & 0.171 & 0.864 & -2.687 & 3.200 \\

\midrule
& & & Coef. & Std.Err. & z & P$>|$z$|$ & [0.025 & 0.975] \\ \cmidrule(lr){3-9}
& & Intercept & 13.214 & 1.254 & 10.536 & 0.000 & 10.756 & 15.672 \\
& & condition[T.B] & -1.539 & 1.444 & -1.066 & 0.286 & -4.370 & 1.291 \\
& & condition[T.C] & 0.711 & 1.414 & 0.503 & 0.615 & -2.061 & 3.482 \\
\multirow{-5}{*}{\centering Learn Quickly} & \multirow{-5}{*}{\parbox{3cm}{\centering I would imagine that most people would learn to use the website very quickly.}} & condition[T.D] & -1.952 & 1.505 & -1.298 & 0.194 & -4.902 & 0.997 \\

\midrule
& & & Coef. & Std.Err. & z & P$>|$z$|$ & [0.025 & 0.975] \\ \cmidrule(lr){3-9}
& & Intercept & 12.424 & 1.261 & 9.849 & 0.000 & 9.951 & 14.896 \\
& & condition[T.B] & -0.717 & 1.505 & -0.476 & 0.634 & -3.666 & 2.233 \\
& & condition[T.C] & 2.078 & 1.462 & 1.421 & 0.155 & -0.787 & 4.944 \\
\multirow{-5}{*}{\centering Intuitive} & \multirow{-5}{*}{\parbox{3cm}{\centering I found the website very intuitive.}} & condition[T.D] & -1.083 & 1.549 & -0.700 & 0.484 & -4.119 & 1.952 \\

\midrule
& & & Coef. & Std.Err. & z & P$>|$z$|$ & [0.025 & 0.975] \\ \cmidrule(lr){3-9}
& & Intercept & 12.624 & 1.273 & 9.918 & 0.000 & 10.130 & 15.119 \\
& & condition[T.B] & -1.245 & 1.543 & -0.807 & 0.420 & -4.270 & 1.780 \\
& & condition[T.C] & 2.034 & 1.493 & 1.362 & 0.173 & -0.892 & 4.960 \\
\multirow{-5}{*}{\centering Confident} & \multirow{-5}{*}{\parbox{3cm}{\centering I felt very confident using the website.}} & condition[T.D] & -1.352 & 1.594 & -0.849 & 0.396 & -4.476 & 1.771 \\

\midrule
& & & Coef. & Std.Err. & z & P$>|$z$|$ & [0.025 & 0.975] \\ \cmidrule(lr){3-9}
& & Intercept & 13.199 & 1.238 & 10.663 & 0.000 & 10.773 & 15.625 \\
& & condition[T.B] & -1.699 & 1.418 & -1.198 & 0.231 & -4.478 & 1.080 \\
& & condition[T.C] & 1.227 & 1.379 & 0.890 & 0.373 & -1.475 & 3.930 \\
\multirow{-5}{*}{No Learning Needed} & \multirow{-5}{*}{\parbox{3cm}{\centering I could use the website without having to learn anything new.}} & condition[T.D] & -2.328 & 1.486 & -1.566 & 0.117 & -5.240 & 0.585 \\
\bottomrule

\end{tabular}
\caption{Aligned rank transform mixed-design ANOVA for system usability scale (SUS). We calculate the aligned rank transform mixed-design ANOVA for each SUS metric, where the values with statistical significance ($\alpha = 0.05$) are highlighted in bold.}
\label{tab:art_anova_sus}
\Description{Statistical analysis of participant post-study responses via ANOVA grouped by user interface settings to show the effect of each setting to system usability.}
\end{table*}

\end{document}